\documentclass{article}
\usepackage{arxiv}

\usepackage{graphics,amssymb,amsmath,epsfig,color}

\usepackage{helvet}
\usepackage{amssymb}
\usepackage{threeparttable}
\usepackage{optidef}
\usepackage[dvipsnames]{xcolor}
\usepackage{amsmath}
\usepackage{array}
\usepackage{multirow}
\usepackage{setspace}

\usepackage{arydshln}
\usepackage{setspace}

\usepackage[sorting=none, backend=biber]{biblatex} 
\DeclareFontFamily{U}{mathb}{\hyphenchar\font45}
\DeclareFontShape{U}{mathb}{m}{n}{
      <5> <6> <7> <8> <9> <10> gen * mathb
      <10.95> mathb10 <12> <14.4> <17.28> <20.74> <24.88> mathb12
      }{}
\DeclareSymbolFont{mathb}{U}{mathb}{m}{n}
\DeclareFontSubstitution{U}{mathb}{m}{n}

\let\dot\relax
\DeclareMathAccent{\dot}{0}{mathb}{"39}
\let\ddot\relax
\DeclareMathAccent{\ddot}{0}{mathb}{"3A}
\let\dddot\relax
\DeclareMathAccent{\dddot}{0}{mathb}{"3B}
\let\ddddot\relax
\DeclareMathAccent{\ddddot}{0}{mathb}{"3C}

\usepackage[english]{babel}
\usepackage[utf8]{inputenc}
\usepackage[T1]{fontenc}
\usepackage{listings}

\usepackage{amsmath}
\usepackage{amssymb}
\usepackage{siunitx}
\usepackage{mathtools}
\usepackage{graphicx}
\usepackage[colorinlistoftodos]{todonotes}
\usepackage{caption}
\usepackage{nameref}
\usepackage[bottom]{footmisc}
\usepackage{bm}

\usepackage{color}
\usepackage[english]{babel}
\usepackage[utf8]{inputenc}
\usepackage[T1]{fontenc}
\usepackage{listings}

\usepackage{amsmath}
\usepackage{amssymb}
\usepackage{siunitx}
\usepackage{mathtools}
\usepackage{graphicx}
\usepackage[colorinlistoftodos]{todonotes}
\usepackage[colorlinks=true, allcolors=black]{hyperref}
\usepackage{caption}
\usepackage[bottom]{footmisc}
\usepackage{bm}
\usepackage{arydshln}
\usepackage{setspace}
\usepackage{multirow}
\usepackage{comment}

\usepackage{cleveref}

\newcommand{\mx}[1]{\begin{bmatrix} #1 \end{bmatrix}}
\newcommand{\eqal}[1]{\begin{equation} \begin{aligned} #1 \end{aligned} \end{equation}}
\newcommand{\eq}[1]{\begin{equation} #1 \end{equation}}

\newcommand{\bnorm}[1]{\left\lVert#1\right\rVert}
\newcommand{\bigb}[1]{\left( #1 \right)}
\newcommand{\tx}[1]{\bm{\tilde{#1}}}
\newcommand{\mino}[1]{\underset{\substack{#1}}{\text{min}}}

\newcommand{\minot}[1]{\underset{\substack{#1}}{\text{minimize}}}

\newcommand{\argmin}[1]{\underset{\substack{#1}}{\text{argmin}}}

\newcommand*{\revhl}{\textcolor{Black}}

\DeclarePairedDelimiter{\abs}{\lvert}{\rvert}

\usepackage{xr}

\makeatletter
\def\endthebibliography{%
  \def\@noitemerr{\@latex@warning{Empty `thebibliography' environment}}%
  \endlist
}
\makeatother

\title{\Large Constrained Ellipse Fitting for Efficient Parameter Mapping with Phase-cycled bSSFP MRI}
\author{Kübra~Keskin\textsuperscript{1}, Uğur~Yılmaz\textsuperscript{2}, Tolga~\c{C}ukur\textsuperscript{2,3} \thanks{This work was supported in part by a European Molecular Biology Organization Installation Grant (IG 3028), by a TUBA GEBIP 2015 fellowship, by a BAGEP 2017 fellowship, and by a TUBITAK 1001 Research Grant (117E171).}
\And
\\
\textsuperscript{1} Department of Electrical and Computer Engineering, University of Southern California, Los Angeles, CA 90089, USA\\
\textsuperscript{2} National Magnetic Resonance Research Center, Bilkent University, Ankara 06800, Turkey\\
\textsuperscript{3} Department of Electrical and Electronics Engineering, Bilkent University, Ankara 06800, Turkey\\
}

\renewcommand{\shorttitle}{Constrained Ellipse Fitting for Parameter Mapping}

\begin{document}

%

\maketitle

\begin{abstract}
Balanced steady-state free precession (bSSFP) imaging enables high scan efficiency in MRI, but differs from conventional sequences in terms of elevated sensitivity to main field inhomogeneity and nonstandard $T_2$/$T_1$-weighted tissue contrast. 
To address these limitations, multiple bSSFP images of the same anatomy are commonly acquired with a set of different RF phase-cycling increments. Joint processing of phase-cycled acquisitions serves to mitigate sensitivity to field inhomogeneity. 
Recently phase-cycled bSSFP acquisitions were also leveraged to estimate relaxation parameters based on explicit signal models. 
While effective, these model-based methods often involve a large number of acquisitions (N$\approx$10-16), degrading scan efficiency.
Here, we propose a new constrained ellipse fitting method (CELF) for parameter estimation with improved efficiency and accuracy in phase-cycled bSSFP MRI. 
CELF is based on the elliptical signal model framework for complex bSSFP signals; and it introduces geometrical constraints on ellipse properties to improve estimation efficiency, and dictionary-based identification to improve estimation accuracy. 
\revhl{CELF generates maps of $T_1$, $T_2$, off-resonance and on-resonant bSSFP signal by employing a separate $B_1$ map to mitigate sensitivity to flip angle variations. Our results indicate that CELF can produce accurate off-resonance and banding-free bSSFP maps with as few as N=4 acquisitions, while estimation accuracy for relaxation parameters is notably limited by biases from microstructural sensitivity of bSSFP imaging.} 
\end{abstract}


\begin{refsection}

\section{Introduction}

Balanced steady-state free precession (bSSFP) is an MRI sequence that offers very high signal-to-noise ratios (SNR) in short scan times \cite{bernstein2004handbook}. Yet this efficiency comes at the expense of distinct signal characteristics compared to conventional sequences. A main difference is the sensitivity of the bSSFP signal to main field inhomogeneities, which causes banding artifacts near regions of large off-resonance shifts \cite{scheffler2003principles}. While a number of approaches were proposed to mitigate artifacts \cite{lee2009improved,benkert2015dynamically,slawig2017multifrequency,kim2017artificial,roeloffs2019frequency}, arguably the most common method is phase-cycled bSSFP imaging \cite{bangerter2004analysis}. Multiple acquisitions of the same anatomy are first collected for a range of different RF phase-cycling increments. A signal combination across phase cycles then reduces severity of artifacts \cite{bangerter2004analysis,ccukur2007enhanced,elliott2007nonlinear,ccukur2008multiple}. Intensity-based combinations assume that banding artifacts are spatially non-overlapping across phase cycles, and that for each voxel there is at least one phase cycle with high signal intensity. In practice, intensity-based methods can perform suboptimally for certain ranges of relaxation parameters or flip angles \cite{bangerter2004analysis}. For improved artifact suppression, recent studies have proposed model-based approaches to estimate the banding-free component of the bSSFP signal. To do this, Linearization for Off-Resonance Estimation-Gauss Newton (LORE-GN) uses a linearized model of the nonlinear bSSFP signal \cite{bjork2014parameter}; Elliptical Signal Model (ESM) employs an elliptical model for the complex bSSFP signal \cite{xiang2014banding,hoff2017combined}; and dictionary-based methods such as trueCISS simulate the bSSFP signal for various tissue/imaging parameters and then perform identification via comparisons against actual measurements \cite{hilbert2018true}. These model-based techniques can perform reliably across practical ranges of tissue and sequence parameters.

An equally important difference of bSSFP is its nonstandard ${T_2}/{T_1}$-dependent contrast \cite{hargreaves2012rapid,bieri2013fundamentals}, which can be considered an advantage for applications focusing on bright fluid signals such as cardiac imaging and angiography \cite{bangerter2011three,ccukur2011magnetization,ilicak2016targeted,deshpande20013d}. That said, bSSFP might yield suboptimal contrast for other applications that focus on primarily $T_2$- or $T_1$-driven differences among soft tissues. A powerful approach to address this limitation is to estimate relaxation parameters from bSSFP acquisitions, and to either directly examine the estimates \cite{margaret2012practical} or synthesize images of desired contrasts \cite{kim2018multicontrast}. Many previous methods in this domain rely on introduction of additional $T_1$- or $T_2$-weighting information into measurements. Given prior knowledge of $T_1$ values, Driven-equilibrium single pulse observation of $T_2$ (DESPOT2) estimates $T_2$ values from N=2-4 acquisitions using a linearized signal equation \cite{deoni2003rapid,deoni2009transverse}. Yet, $T_1$ values are typically estimated via two SPGR acquisitions at different flip angles that are not as scan efficient as bSSFP. An alternative is to use magnetization-preparation modules in conjunction with bSSFP to enable $T_1$ estimation \cite{scheffler2001t1} or simulatenous $T_1$ and $T_2$ estimation \cite{schmitt2004inversion,ehses2013ir}. Magnetization-preparation modules reduce scan efficiency, and these methods can be susceptible to field inhomogeneity when based on a single phase-cycled acquisition. Dictionary-based methods for parameter estimation have also been combined with bSSFP sequences with varying parameters across readouts as in Magnetic Resonance Fingerprinting \cite{ma2013magnetic}. While dictionary-based methods enable estimation of many tissue and imaging parameters including $T_1$ and $T_2$, they require advanced pulse sequence designs that may not be available at all sites. 

A promising approach based on standard bSSFP sequences is artificial neural networks for purely data-driven estimation of $T_1$ and $T_2$ maps. This learning-based method requires training data from a lengthy protocol with bSSFP and gold-standard mapping sequences; and it might require retraining for different protocols or pathologies \cite{heule2020multi}. In contrast, the model-based trueCISS method uses N=16 acquisitions to compare the measured signal profile across phase cycles against a simulated dictionary of signal profiles \cite{hilbert2018true}. While trueCISS can estimate the equilibrium magnetization, the $T_1/T_2$ ratio and off-resonance, it does not consider explicit estimation of $T_1$ or $T_2$. LORE-GN instead performs least-squares estimation of $T_1$ and $T_2$ based on linearized equations \cite{bjork2014parameter}. That said, accurate $T_1$ and $T_2$ estimation was suggested to be difficult with N=4 at practical SNR levels. MIRACLE also uses multiple bSSFP acquisitions to perform configuration based TESS relaxometry where N=8-12 \cite{nguyen2017motion}. Recently, the PLANET method based on ESM was proposed to demonstrate feasability of simultaneous $T_1$ and $T_2$ estimation \cite{shcherbakova2018planet}. PLANET first fits an ellipse to multiple-acquisition bSSFP measurements, and then analytically estimates $T_1$ and $T_2$ values for each voxel by observing geometric properties \cite{shcherbakova2018planet}. Theoretical treatment suggested that N$\ge$6 is required for ellipse fitting; and demonstrations were performed at N=10 to improve noise resilience \cite{shcherbakova2018planet,shcherbakova2019accuracy}. Despite the elegance of these recent approaches, the use of relatively large N partly limits scan efficiency. 

Here, we propose a new method, CELF, for parameter estimation with fewer number of bSSFP acquisitions, structured upon the ESM framework. To improve efficiency, CELF introduces additional geometric constraints to ellipse fitting, such that the number of unknowns and the required number of acquisitions can be reduced to N=4. The central line of the ellipse is identified via a geometric solution \cite{xiang2014banding}, and then incorporated as prior knowledge for a constrained ellipse fit. To improve accuracy, a dictionary-based identification is introduced to obtain a final ellipse fit. $T_1$, $T_2$, off-resonance, and on-resonant signal intensity values are analytically derived from the geometric properties of the ellipse. To mitigate sensitivity to flip angle variations, CELF employs a separately acquired $B_1$ map. Comprehensive evaluations are performed via simulations as well as phantom and in vivo experiments. 
Our results suggest that CELF yields improved efficiency in parameter estimation compared to direct ellipse fitting, and it can produce accurate off-resonance and banding-free bSSFP maps with as few as N=4 acquisitions. Meanwhile, in vivo estimates of relaxation parameters carry biases from microstructural sensitivity of bSSFP that limit estimation accuracy compared to conventional spin-echo methods, so future mitigation efforts for these biases will be essential to ensure practicality of CELF in relaxometry applications. 

\subsubsection*{\textbf{Contributions}}
The novel contributions of this study are summarized below:
\begin{itemize}
    \item We introduce an ellipse fitting procedure with a geometric constraint on the ellipse's central line for estimation of $T_1$, $T_2$, off-resonance, and on-resonant signal values from phase-cycled bSSFP MRI with as few as N=4 acquisitions.
    \item We derive an analytical solution for the constrained ellipse fit to avoid brute-force search or iterative optimization approaches. 
     \item We introduce a dictionary-based identification procedure on the ellipse fits to further improve estimation accuracy. Identification is performed on the ellipses as opposed to raw bSSFP signals for reliability against off-resonance. 
\end{itemize}

\section{Theory and Methods}
The proposed method leverages an elliptical signal model for the bSSFP signal to estimate the equilibrium magnetization, off-resonance, and $T_1$ and $T_2$ values. In contrast to direct ellipse fitting, CELF employs additional geometric constraints to improve scan efficiency and a dictionary-based ellipse identification to improve accuracy. In the following subsections, we overview ESM and direct ellipse fitting. We then describe the proposed constrained ellipse fitting framework.

\subsection{Elliptical Signal Model for Phase-cycled bSSFP}

\subsubsection*{\textbf{Analytical expression of the bSSFP signal}}
The steady-state signal generated by a phase-cycled bSSFP sequence right after the radio-frequency (RF) excitation can be expressed as \cite{xiang2014banding}:
\begin{equation} \label{eq:signal_model}
S_{base}(r) = M(r) \frac{ 1-a(r) e^{i\theta(r)}}{1-b(r)\cos{\theta(r)}}
\end{equation}
where
\begin{equation} \label{eq:signal_params}
\begin{aligned}
a(r) = E_2(r) , \; b(r) = \tfrac{ E_2(r)(1-E_1(r))(1+\cos{\alpha(r)})}{1-E_1(r)\cos{\alpha(r)}-E_2(r)^2(E_1(r)-\cos{\alpha(r)})} \\
M(r) = \tfrac{M_0(r)(1-E_1(r))\sin{\alpha(r)}}{1-E_1(r)\cos{\alpha(r)}-E_2(r)^2(E_1(r)-\cos{\alpha(r)})}
\end{aligned}
\end{equation}
In Eq.~\eqref{eq:signal_params}, $r$ denotes spatial location, $E_1(r)=exp(-TR/T_1(r))$ and $E_2(r)=exp(-TR/T_2(r))$ characterize exponential decay for longitudinal and transverse magnetization with $T_1$ and $T_2$  relaxation times. TR denotes the repetition time, $M_0$ is the equilibrium magnetization, and $\alpha$ is the spatially-varying flip angle of the RF excitation. Meanwhile, $\theta = \theta_0 - \Delta\theta$, where $\theta_0 = 2\pi (\Delta f_0+\delta_{cs})$TR reflects phase accrual due to main field inhomogeneity at off-resonance frequency $\Delta f_0$ and chemical shift frequency $\delta_{cs}$, and $\Delta\theta$ reflects phase accrual due to RF phase-cycling. In multiple-acquisition bSSFP, N separate measurements are obtained while the $\Delta\theta$ is spanned across $[0,2\pi)$ in equispaced intervals (e.g. $\Delta\theta=\{0,\pi/2,\pi,3\pi/2\}$ for N=4).

For measurements performed at the echo time TE, the base signal expressed in Eq.~\eqref{eq:signal_model} gets scaled and rotated as follows:
\begin{equation} \label{eq:rot_signal}
S(r) = K(r)M(r) e^{-TE/T_{2}(r)} \frac{ 1-a(r) e^{i\theta(r)}}{1-b(r)\cos{\theta(r)}} e^{i\phi(r)}
\end{equation}
where K is a complex scalar denoting coil sensitivity, $\phi = 2\pi (\Delta f_0+\delta_{cs}) TE + \phi_{total}$ represents the aggregate phase accrual, and $\phi_{total}$ captures phase accumulation due to system imperfections including eddy currents and main field drifts.

The unknown tissue-dependent parameters to be estimated are $T_1$, $T_2$, $M_0$ and $\Delta f_0$. In this study, based on preliminary observations, we assume that phase accrued due to eddy currents and main field drifts is stationary across separate phase cycles, so it reflects a constant phase offset for all acquisitions. Meanwhile, the values of user-controlled imaging parameters $TR$, $TE$ and $\Delta\theta$ are known. To mitigate RF field inhomogeneity, we include a $B_1$-mapping scan in the MRI protocol for this study, so we assume that a spatial map of the flip angle $\alpha$ will be available. Thus, $M$, $a$, $b$ and $\theta$ that parametrize the bSSFP signal in Eq.~\eqref{eq:rot_signal} depend on a priori known or measured parameters via nonlinear relations. 

\subsubsection*{\textbf{Elliptical Signal Model}}
\begin{figure}[!ht]
\centering
\includegraphics[width=1\columnwidth]{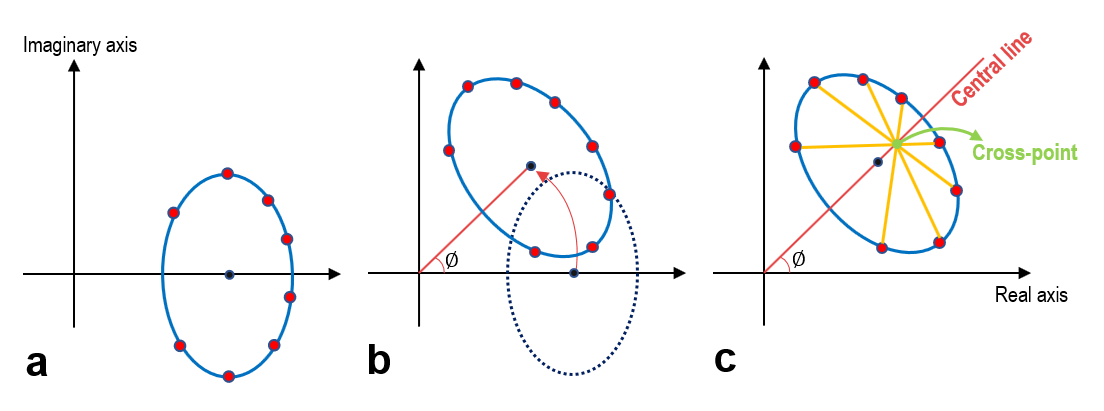}

\caption{\label{fig:ellipse} (a) The complex-valued base bSSFP signal $S_{base}$ following the RF excitation defines a vertically-oriented ellipse. The signal values at different phase-cycling increments $\Delta\theta$ project onto separate points on this ellipse (shown with red dots). (b) The measured bSSFP signal at echo time (TE) constitutes a scaled and rotated version of the ellipse for $S_{base}$. (c) The ellipse cross-point is at the intersection of line segments that connect measurement pairs: $(i,j)$ is a pair if $|\Delta\theta_i-\Delta\theta_j|=\pi$. The central line of the ellipse passes through both the ellipse center and cross-point.}

\end{figure}

A powerful framework to examine the link between the signal parameters and the actual measurements is the elliptical signal model (ESM) by Xiang and Hoff \cite{xiang2014banding}. The ESM framework observes that for a given voxel Eq.~\eqref{eq:signal_model} describes an ellipse in the complex plane for bSSFP signals. Each bSSFP measurement acquired with a specific phase-cycling increment ($\Delta\theta$) projects onto a point on this ellipse, as demonstrated in Fig. \ref{fig:ellipse}a. Characteristic properties of the bSSFP ellipse in terms of the parameters in Eq.~\eqref{eq:signal_params} are defined below \cite{xiang2014banding}:

\begin{itemize}  \label{itm:prop_ell}
    \item Semi-major, semi-minor axes: $M\frac{a}{\sqrt{1-b^2}}$ and $M\frac{|a-b|}{1-b^2}$ \hfill$\refstepcounter{equation}(\theequation)\label{eq:esm_item1}$
    \item Geometric center: $(M\frac{1-ab}{1-b^2},0)$\hfill$\refstepcounter{equation}(\theequation)\label{eq:esm_item2}$
    \item Eccentricity: $\sqrt{1-\frac{(a-b)^2}{a^2(1-b^2)}}$\hfill$\refstepcounter{equation}(\theequation)\label{eq:esm_item3}$
    \item Condition for vertically-oriented ellipse: $b<\frac{2a}{1+a^2}$\hfill$\refstepcounter{equation}(\theequation)\label{eq:esm_item4}$
\end{itemize}

Note that the ellipse for the measured bSSFP signals in Eq.~\eqref{eq:rot_signal} is a scaled and rotated version of the ellipse of the base signals in Eq.~\eqref{eq:signal_model}, as illustrated in Fig. \ref{fig:ellipse}b.

\subsubsection*{\textbf{Geometric Solution}}
An effective method to analytically compute an on-resonant bSSFP image via ESM involves the geometric solution (GS) of the ellipse \cite{xiang2014banding}. Measurements are paired based on the difference between phase-cycling increments such that $(i,j)$ is a pair if $|\Delta\theta_i-\Delta\theta_j|=\pi$. GS is defined as the cross-point of all formulated pairs. For example, the pairs for N=4 are: {(1,3) with $|0-\pi| = \pi$; (2,4) with $|\frac{\pi}{2}-\frac{3\pi}{2}| = \pi$}. Note that GS of the ellipse in Eq.~\eqref{eq:signal_model} corresponds to the parameter $M$ in Eq.~\eqref{eq:signal_params}, and GS of the ellipse in Eq.~\eqref{eq:rot_signal} is a scaled and rotated version of $M$. Since $M$ does not depend on $\theta$, it represents an on-resonant bSSFP image \cite{xiang2014banding}. The original GS can be extended to N acquisitions as follows:
\begin{equation} \label{eq:cross_point}
\mx{ 
y_{m+1}-y_{1} & x_{1}-x_{m+1}\\
\vdots & \vdots \\
y_{N}-y_{m} & x_{m}-x_{N}
}
\mx{x_0\\y_0}
=
\mx{
x_{1}y_{m+1}-x_{m+1}y_{1}\\
\vdots\\
x_{m}y_{N}-x_{N}y_{m}
}
\end{equation}

where $(x,y)_i$ denote the real and imaginary components of $S$ in the $i^{th}$ acquisition, $(x,y)_{i}$ and $(x,y)_{m+i}$ form a $\pi$-separated signal pair ($m=N/2$ and $i={1,2,..,m}$), and $q = (x_0,y_0)$ is the cross-point that can be estimated by ordinary least-squares.

\subsection{Direct Ellipse Fitting}
The unknown parameters ($T_1$, $T_2$, $M_0$, $\Delta f_0$) can be estimated by fitting an ellipse to the collection of N phase-cycled bSSFP signals for a given voxel. To do this, the PLANET method uses Fitzgibbon's direct least square fitting of ellipses \cite{fitzgibbon1999direct}. In quadratic form, an ellipse can be described as:
\begin{equation} \label{eq:def_fun}
    f(\bm{x}) = \nu_1 x^2 + \nu_2 xy + \nu_3 y^2 + \nu_4 x + \nu_5 y + \nu_6 = 0
\end{equation}
where $\bm{\nu}=\mx{\nu_{1:6}}^T$ is the polynomial coefficient vector, and $\bm{x}=\mx{x & y}^T$ is a point on the ellipse where $x$ and $y$ denote the real and imaginary components of $S$. Minimizing algebraic distance between measured data points and fit ellipse, the following least-squares formulation can be obtained: 
\begin{equation} \label{eq:defCost} 
\begin{aligned}
     \bnorm{\mx{ f(\bm{x}_1) \\ \vdots \\ f(\bm{x}_N)}}^2_2 
     &= \bnorm{\mx{x_1^2 & x_1y_1 & y_1^2 & x_1 & y_1 & 1 \\
    \vdots & \vdots & \vdots & \vdots & \vdots & \vdots \\
    x_N^2 & x_Ny_N & y_N^2 & x_N & y_N & 1} \mx{\nu_1\\ \vdots \\ \nu_6} }^2_2 \\
    &= \bnorm{D\bm{\nu}}^2_2
\end{aligned}
\end{equation} 
where $f(\bm{x}_i)$ denotes the value of the polynomial function evaluated at the $i^{th}$ data point $\bm{x}_i$, and $D$ is the aggregate data matrix for N acquisitions. Minimization of sum of squared errors $\sum_{i=1}^N{|f(\bm{x}_i)|^2}$ yields:
\begin{equation} \label{eq:direct_fitting}
\begin{aligned}
& \minot{\bm{\nu}} & & \bnorm{D\bm{\nu}}^2_2 \\ 
& \text{subject to} & & \bm{\nu}^T C \bm{\nu} = 1
\end{aligned}
\end{equation}
where 
$C = \left[
    \begin{array}{c;{2pt/2pt}r}
        \begin{matrix} 0&0&2\\0&-1&0\\2&0&0 \end{matrix} & 0_{3\times3} \\ \hdashline[2pt/2pt]
        0_{3\times3}   & 0_{3\times3}
    \end{array}
\right]$ is the constraint matrix ensuring that the fit polynomial is an ellipse. An analytical solution can be obtained by solving an equivalent generalized eigenvalue problem (GEP):
\begin{equation}
\label{eq:geneig}
    D^TD\bm{\nu} = \lambda C \bm{\nu}
\end{equation}

In sum, Fitzgibbon's method solves Eq.~\eqref{eq:geneig} to obtain the best fit ellipse in a non-iterative and numerically stable way \cite{halir1998numerically}. The quadratic form of an ellipse is uniquely specified by six scalar coefficients in $\bm{\nu}$, so a theoretical minimum of N=6 is needed for fitting. Yet MRI acquisitions are inherently noisy, and this can reduce fit accuracy for relatively lower N. To mitigate this problem, previous studies have used up to N=10 at the expense of prolonged scan times \cite{shcherbakova2018planet,shcherbakova2019accuracy}.
In this paper, PLANET was used for direct ellipse fitting as described in \cite{shcherbakova2018planet}. In PLANET, the ellipse for $S$ is back-rotated to vertical orientation with a rotation angle $\varphi_{rot}=0.5\tan^{-1}\bigb{\nu_2/(\nu_1-\nu_3)}$. Here, we observed that $\varphi_{rot}$ does not always assure a strict vertical orientation due to noise. We reasoned that the rotation angle that back-rotates the fit ellipse should be $\phi$. Therefore, we implemented two variants: back-rotation with $\varphi_{rot}$, and with $\phi$. Since we observed that the latter is less prone to estimation errors, here we presented results from back-rotation with $\phi$ variant. Comparison of the two variants is given in Supp. Fig. 1.

\subsection{Constrained Ellipse Fitting}
To improve scan efficiency, constrained ellipse fitting (CELF) incorporates geometric prior knowledge to enable unique ellipse specification with N=4 \cite{waibel2015constrained}. Specifically, the ellipse center and thereby the ellipse orientation are identified via the geometric solution. (Note that in recent studies the GS phase was also used for correcting geometric distortions \cite{NicholasHoff2010} and mapping the main field strength \cite{Taylor2017}.) To improve fit quality, CELF then employs dictionary-based ellipse identification. Parameters are extracted from the identified ellipses. These steps are described below.

\subsubsection*{\textbf{Formulation of CELF}}
To facilitate integration of the geometric prior, here we use the matrix form of a linearly translated central conic quadratic equation to describe the ellipse:
\begin{equation} \label{eq:f1_fun}
    f(\bm{x}) = (\bm{x}-\bm{x_c})^T A (\bm{x}-\bm{x_c}) + g = 0 
\end{equation}
where $A = \mx{c_1 & \frac{c_2}{2}\\\frac{c_2}{2} & c_3}$ is a positive definite matrix describing the ellipse orientation and eccentricity, $\bm{x}=\mx{x & y}^T$ is a point on the ellipse, $\bm{x_c}=\mx{x_c & y_c}^T$ is the ellipse center, and g is a finite negative scalar determining the ellipse area. The scalar form of Eq.~\eqref{eq:f1_fun} is given as:
\begin{equation} \label{eq:f1_funs}
c_1(x-x_c)^2 + c_2(x-x_c)(y-y_c) + c_3(y-y_c)^2 + g = 0
\end{equation}
Eq.~\eqref{eq:f1_funs} still has six unknowns: $c_1$, $c_2$, $c_3$, $g$, $x_c$ and $y_c$.

CELF uses prior knowledge on the ellipse's central line for improved ellipse fitting at lower N. To leverage this prior, we first observe that the base bSSFP signal $S_{base}$ in Eq.~\eqref{eq:signal_model} forms a vertically-oriented ellipse ---with the semi-major axis perpendicular to the line passing through its center and the origin (see Fig. \ref{fig:ellipse}a)---, whereas the measured bSSFP signal $S$ in Eq.~\eqref{eq:rot_signal} can be obtained by rotating the base ellipse at an arbitrary angle (see Fig. \ref{fig:ellipse}b). It can be also observed that both the ellipse center and the cross-point lie on the semi-minor axis. Thus, the semi-minor axis is a line segment on the central line, and the central line can be identified by finding the cross-point (see Fig. \ref{fig:ellipse}c). Here we found the cross-point $q$ via the geometric solution described in Eq.~\eqref{eq:cross_point}. Therefore, to produce the vertically-oriented base version, the ellipse for $S$ can be back-rotated around the origin by an angle $\phi$, i.e. the angle that the central line makes with the real axis. (Note that the vertical-orientation constraint in CELF ensures that $\varphi_{rot}$ equals $\phi$.) This back-rotation will place the central line of the ellipse along the x-axis. The orientation-eccentricity matrix will then be diagonal ($c_2=0$; $\tilde{A}=\mx{c_1&0\\0&c_3}$), the center $\bm{\tilde{x}_c}$ will only have a real component. 

Next, we express the ellipse center as a scaled version of the cross-point $q$, which is given by the geometric solution. In particular, $\bm{\tilde{x}_c}= \mx{x_c & y_c} = \mx{\gamma q & 0}^T$, where $\gamma$ is an unknown scaling parameter. With a transformation of the coordinate system, the spatial coordinates of the data points in the back-rotated ellipse can be described as $\bm{\tilde{x}}=\mx{\tilde{x} & \tilde{y}}^T$. The ellipse equation in Eq.~\eqref{eq:f1_fun} then becomes:
\begin{equation}
\begin{aligned}
    f(\bm{\tilde{x}}) &= \mx{ \tilde{x}-\gamma q & \tilde{y} } \mx{c_1&0\\0&c_3} \mx{ \tilde{x}-\gamma q \\ \tilde{y} } +g \\
    &= c_1 \tilde{x}^2 + c_3 \tilde{y}^2 -2\gamma c_1 q \tilde{x} + h 
\end{aligned}
\end{equation}
where $h=g+c_1\gamma^2q^2$ is a scalar, and the set of unknowns is reduced to $c_1$, $c_3$, $\gamma$, and $h$. Thus, N=4 acquisitions are sufficient for ellipse fitting in principle.  

Similar to direct ellipse fitting, the error measure of algebraic distance between measured data points and the fit ellipse leads to a least-squares problem:
\begin{equation} \label{eq:costQ}
\begin{aligned}  
     \bnorm{\mx{ f(\bm{\tilde{x}}_1) \\ \vdots \\ f(\bm{\tilde{x}}_N)}}^2_2 
     &=   \bnorm{\bigb{\mx{\tx{x}_1^2 & \tx{y}_1^2 & 1 \\ \vdots & \vdots & \vdots \\ \tx{x}_N^2 & \tx{y}_N^2 & 1 } - 
     \gamma \mx{2q\tx{x}_1 & 0 & 0 \\ \vdots & \vdots & \vdots \\ 2q\tx{x}_N & 0 & 0 } } \mx{c_1 \\ c_3 \\ h} }^2_2 \\
     &=  \bnorm{\bigb{\mx{D_0 & \bm{1}_N} - \gamma \mx{D_1 & \bm{0}_N} } \mx{\bm{u} \\ h} }^2_2 \\
     &= Q(\gamma,\bm{u},h)
\end{aligned}
\end{equation}
where $f(\bm{\tilde{x}}_i)$ denotes the value of the polynomial function evaluated at the $i^{th}$ back-rotated data point $\bm{\tilde{x}}_i$, $D_0$ and $D_1$ are aggregate data matrices for N acquisitions, $\bm{u}=\mx{c_1 & c_3}^T$ is defined as a parameter vector for $\tilde{A}$. The cost function is taken as $Q = \sum_{i=1}^N{|f(\bm{\tilde{x}}_i)|^2}$.

To constrain the solution of Eq.~\eqref{eq:costQ} to a strict ellipse, $A$ must be positive definite, i.e., all leading principal minors of $A$ are positive: $\det\bigb{\mx{c_1}}=c_1>0$ and $\det(A)=(4c_1c_3-c_2^2)/4>0$. To introduce rotation/translation invariance and to reduce the degrees of freedom, $\det(A)$ is set to a constant positive number without loss of generality \cite{fitzgibbon1999direct}. In this case, constraining the fit to an ellipse is equivalent to the following conditions \cite{waibel2015constrained}: $4\det(A)=4c_1c_3-c_2^2=1$ and $c_1>0$. Since $c_2=0$ in the back-rotated form $\tilde{A}$, these conditions simplify to $4c_1c_3=1$ and $c_1>0$. These two conditions are integrated to the ellipse fitting procedure as additional constraints:
\begin{equation}
\begin{aligned}
    \bm{u}^T B \bm{u} &= \mx{c_1 & c_3} \mx{0 & 2 \\ 2 & 0} \mx{c_1 \\ c_3} = 1, \\
    \bm{u}^T\bm{d} &= \mx{c_1&c_3}\mx{1\\0}>0
\end{aligned}
\end{equation}
where $B$ is the constraint matrix and $\bm{d}$ is the constraint vector.

Lastly, defining the cost function in terms of the unknown parameters $\gamma$, $\bm{u}$ and $h$, the constrained ellipse fitting problem can be stated as follows:
\begin{equation} \label{eq:overall_prob}
\begin{aligned}
& \minot{\gamma,\bm{u},h} & & Q(\gamma,\bm{u},h) \\ 
& \text{subject to} & & \bm{u}^T B \bm{u} = 1, \; \text{and } \bm{u}^T\bm{d}>0
\end{aligned}
\end{equation}

\begin{figure*}[th]
\centering
\includegraphics[width=\textwidth]{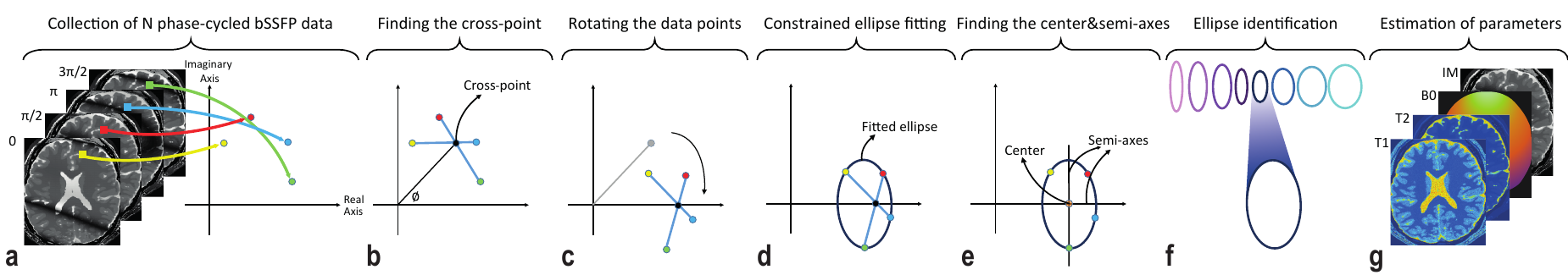}

\caption{\label{fig:steps} Flowchart of the proposed constrained ellipse fitting (CELF) approach. For a given voxel, the ESM model observes that each phase-cycled bSSFP measurement should project to an ellipse in the complex plane. In CELF: (b) The ellipse cross-point is first computed via the geometric solution to identify the central line of the ellipse; (c) Measurement points are back-rotated by the angle between central line and real-axis; (d) The prior knowledge of the cross-point is then used to enable constrained ellipse fitting with as few as N=4; (e) Geometric properties of the fit ellipse including center and semi-axes are extracted; (f) A dictionary-based ellipse identification is performed to improve accuracy; (g) Lastly, parameters estimates are obtained based on the geometric properties of the final ellipse fit.}

\end{figure*}

\subsubsection*{\textbf{Solution of CELF}}
It is challenging to obtain a single-shot solution to Eq.~\eqref{eq:overall_prob} that estimates all unknowns simultaneously as in direct ellipse fitting. Here we instead use a progressive approach to sequentially identify the unknown parameters \cite{waibel2015constrained}:
\begin{equation} \label{eq:successive_prob}
\begin{aligned}
& \mino{\gamma,\bm{u},h \\ \bm{u}^T B \bm{u} = 1 \\ \bm{u}^T\bm{d}>0} 
& Q(\gamma,\bm{u},h) 
&= \mino{\gamma} \; 
& \mino{\bm{u} \\ \bm{u}^T B \bm{u} = 1 \\ \bm{u}^T\bm{d}>0} \; & \mino{h} \; Q(\gamma,\bm{u},h) \\
& & &= \mino{\gamma} \; 
& \mino{\bm{u} \\ \bm{u}^T B \bm{u} = 1 \\ \bm{u}^T\bm{d}>0} \; & Q_2(\gamma,\bm{u}) \\
& & &= \mino{\gamma} \;
& Q_3(\gamma) &
\end{aligned}
\end{equation}
where the minimization is decomposed into three subproblems to estimate $h$, $\bm{u}$, and $\gamma$, respectively.  
  
\textbf{Subproblem 1.} The first subproblem is defined as a minimization over the parameter $h$, and it aims to optimize $h$ conditioned on the values of the remaining parameters ($\gamma, \bm{u}$). The analytical solution to this problem can be expressed as (see Supp. Text 1.1 for derivation):
\begin{equation}
    h_{opt}(\gamma,\bm{u}) = -\frac{1}{N} \bm{1}_N^T(D_0-\gamma D_1)\bm{u}
\end{equation}

\textbf{Subproblem 2.} Once its optimal value is identified, $h$ can be factored out of the optimization by substituting $h_{opt}$ into $Q$. This new cost function is denoted as $Q_2$, and it represents the minimum of $Q$ over $h$ (see Supp. Text 1.2 for derivation):
\begin{equation}
\begin{aligned}
    Q_2(\gamma,\bm{u}) &= \bm{u}^T \bigb{C_0 + \gamma C_1 + \gamma^2 C_2} \bm{u}
\end{aligned}
\end{equation} where $C_0=D_0^T Z D_0$, $C_1=-D_0^T Z D_1 - D_1^T Z D_0$, $C_2=D_1^T Z D_1$ and $Z=I_N-\frac{1}{N}1_{N \times N}$.

The second problem is then defined as a minimization over $\bm{u}$ conditioned on $\gamma$:
\begin{equation} \label{eq:2ndprob}
\begin{aligned}
    & \minot{\bm{u}} & & \bm{u}^T \bigb{C_0 + \gamma C_1 + \gamma^2 C_2} \bm{u} \\ 
    & \text{subject to} & & \bm{u}^T B \bm{u} = 1, \; \text{and } \bm{u}^T\bm{d}>0
\end{aligned}
\end{equation} 
This subproblem can be cast as a generalized eigenvalue problem (see Supp. Text 1.3 for proof):
\begin{equation} \label{eq:gep}
    \bigb{C_0 + \gamma C_1 + \gamma^2 C_2} \bm{u} = \lambda B \bm{u}
\end{equation}
The minimum value of the function $Q_2$ is the maximum eigenvalue $\lambda_{max}$, and this minimum is attained at the respective eigenvector $\bm{u}_{opt}$ (see Supp. 1.3 for derivation):
\begin{equation}
    \bm{u}_{opt}(\gamma) = \frac{\text{sign}(\bm{u}_{max}^T\bm{d})}{\sqrt{\bm{u}_{max}^T B \bm{u}_{max}}} \bm{u}_{max}
\end{equation}

\textbf{Subproblem 3.} Once $\bm{u}_{opt}$ is identified, it is factored out by substitution into $Q_2$. This results in a new cost function denoted as $Q_3$ depending only on $\gamma$ that controls the location of the ellipse center (see Supp. Text 1.4 for detailed derivation):
\begin{equation}
    Q_3(\gamma) = \lambda_{max}\bigb{C_0 + \gamma C_1 + \gamma^2 C_2,B}
\end{equation}
Thus, Eq.~\eqref{eq:overall_prob} is equivalent to:
\begin{equation} \label{eq:min_q3}
    \minot{\gamma} \; Q_3(\gamma)
\end{equation}

In sum, the original multi-dimensional minimization is reduced to a one-dimensional problem, where we seek $\gamma$ that minimizes $\lambda_{max}$. Such problems are commonly solved via iterative methods or brute-force search over $\gamma$. For computational efficiency, however, we derive for the first time in this study the analytical solution for $\gamma$ (see Supp. Text 1.5 for derivation), which disregards measurement noise. Yet, excessive noise can cause complex valued $\gamma_{1,2}$ (Eq.~(S.15)). In such cases, we instead selected $\gamma$ through a bounded search to minimize $\lambda_{max}$. Note that the ellipse center given by $x_c = M\frac{1-ab}{1-b^2}$ can also be expressed in terms of the cross-point as $x_c = \gamma q$. Since $q$ corresponds to $M$, $\gamma$ should ideally equal $\frac{1-ab}{1-b^2}$. We computed $\gamma$ for all possible signal parameters $a$ and $b$ where $T_1$ $\in $ [200 5000] ms, $T_2$ $\in$ [10 1500] ms, flip angle $\in$ [20 80]$^o$, and TR $\in$ [4 10] ms (see Supp. Fig. 2 for an example where TR is 8 ms, and flip angle is $40^o$). The bounded search was accordingly restricted to $\gamma$ $\in$ [0.5 1]. Once the optimum value is identified, the remaining parameters can be extracted from $\gamma^*$ as follows:
\begin{equation} \label{eq:param_sols}
\begin{aligned}
    \bm{u}^* &= \bm{u}_{opt}(\gamma^*) \\
    h^* &= h_{opt}(\gamma^*,\bm{u}^*) \\
    A^* &= \mx{u_1^* & 0 \\ 0 & u_2^*} \\
    \bm{x}_c^* &= \mx{\gamma^* q & 0} \\
    g^* &= h^* - \gamma^{*2} q^2 \bm{u}^{*T}\bm{d}
\end{aligned}
\end{equation}

The constrained formulation allows for an ellipse fit with as few as N=4 samples. Note that, CELF estimates have several singularities precisely localized to $\theta_0=\{\pm\frac{\pi}{4},\pm\frac{3\pi}{4}\}$ for N=4. For these $\theta_0$, bSSFP signals on the vertically-oriented ellipse are symmetrically distributed about the x-axis, forming two pairs of conjugate symmetric data points (see Supp. Fig. 3). As such, only 2 independent measurements are available to solve \eqref{eq:costQ}, leading to an underdetermined system. While the singularities do not affect banding-free signal or off-resonance estimates, they might introduce errors in $T_1$ and $T_2$ estimates. To address this issue, a voxel-wise singularity detection was performed (see Supp. Text 1.6 for details). For a singular voxel, data from the immediate $3\times3$ neighborhood of the central voxel within axial cross-sections were aggregated, and the ellipse fit was performed based on the aggregate data. Note that all processing stages in CELF operate on single voxels, with the exception of singular voxels.

\subsubsection*{\textbf{Dictionary-based ellipse identification}}
   Inherent noise in bSSFP measurements can limit the accuracy of ellipse fits with lower N. This in turn can degrade the quality of $T_1$ and $T_2$ estimates. To improve ellipse fits, here we introduce a dictionary-based ellipse identification procedure. The dictionary only contains the following ellipse properties: the semi-axes radii and the distance between the ellipse center and the origin. Since these properties are not affected by main-field inhomogeneity, we did not consider off-resonance. The simulations also excluded non-stationary effects on bSSFP signals due to motion, eddy currents, drifts and noise. 
   
To construct the dictionary, we simulated bSSFP signal parameters $M$, $a$ and $b$ according to Eq.~\eqref{eq:signal_params}. The simulations used $TR$, $TE$, and nominal flip angle $\alpha$ matched to the bSSFP sequence whose measurements are to be analyzed. Ellipses were simulated for all possible pairs ($T_1$,$T_2$), where $T_1$ ranged from 50 to 5000 ms in 5-ms steps, and $T_2$ ranged from 10 to 500 ms in 1-ms steps and from 500 to 1500 ms in 5-ms steps. Simulated $M$, $a$, $b$ values were then used to calculate the semi-axes radii and the distance between the ellipse center and the origin (Eqs.~\eqref{eq:esm_item1}-\eqref{eq:esm_item2}). The equilibrium magnetization was taken as 1 during the simulations without loss of generality, since ellipse properties were normalized by the cross-point distances (i.e. $|M|$ for the dictionary ellipse). The resulting dictionary contains semi-axes radii and center distance properties for all pairs of $T_1$-$T_2$ examined.

   For identification, a comparison is performed between properties of the dictionary ellipses and the fit ellipse for $S$. The semi-axes radii of the fit ellipse can be extracted from the solution in Eq.~\eqref{eq:param_sols} as follows:
\begin{itemize}  \label{itm:cprop_ell}
    \item Semi-minor radius: $r_{min} = \sqrt{-{g^*}/{u_1^*}}$\hfill$\refstepcounter{equation}(\theequation)\label{eq:cesm_item2}$
    \item Semi-major radius: $r_{maj} = \sqrt{-{g^*}/{u_2^*}}$\hfill$\refstepcounter{equation}(\theequation)\label{eq:cesm_item4}$
\end{itemize}   
Meanwhile, the center distance can be directly calculated from the $x_c^*$ in Eq.~\eqref{eq:param_sols}. The summed $\ell_2-norm$ distance between the properties of the dictionary and fit ellipse's are computed. To prevent biases due to differences in signal scales, all ellipse properties were normalized by the cross-point distances (i.e. $|M|$ for the dictionary ellipse, $|q|$ for the fit ellipse). Distances between the dictionary and fit ellipses were computed following this normalization. The dictionary ellipse that minimizes the distance to the fit ellipse was then selected as the final ellipse estimate.

\subsubsection*{\textbf{Parameter estimation}}   
   Given the center $x_c^*$ and semi-axes radii $r_{min}$ and $r_{maj}$ of the final ellipse estimate, we can compute the parameters in Eq.~\eqref{eq:signal_params} as follows \cite{shcherbakova2018planet}:
    \eqal{         
        b^* &= \frac{-r_{min}x_c^*+r_{maj}\sqrt{x_c^{*2}-r_{min}^2+r_{maj}^2}}{x_c^{*2}+r_{maj}^2} \\
         a^* &= \frac{r_{maj}}{x_c^*\sqrt{1-b^{*2}}+r_{maj}b^*}, \; M^* = \frac{x_c^*(1-b^{*2})}{1-a^*b^*}
    }

Once ($M^*$, $a^*$, $b^*$) are known, $T_1$ and $T_2$ values can be derived analytically \cite{shcherbakova2018planet}:
    \eq{ \label{eq:T1T2Est}
         T_1 = -\frac{TR}{\ln{\frac{a^*(1+\cos{\alpha}-a^*b^*\cos{\alpha})-b^*}{a^*(1+\cos{\alpha}-a^*b^*)-b^*\cos{\alpha}}}}, \; T_2 = -\frac{TR}{\ln{a^*}}
    }
    
Please note that ellipse fits do not yield estimates of the flip angle. Since the actual flip angle for a voxel might deviate from the nominal flip angle prescribed during the bSSFP scans, $\alpha$ in Eq.~\eqref{eq:T1T2Est} is taken as the measured flip angle corrected according to the $B_1$ mapping scan. Note that $M^*$ serves as an estimate for the banding-free image as it does not show any off-resonance dependency \cite{xiang2014banding}. Meanwhile, given ($\tilde{x}_i$, $x_c^*$, $r_{min}$, $b^*$, $\Delta\theta_i$), off-resonance in each voxel can be calculated via ordinary least squares solution of the following system of equations:
\eq{
\mx{
\cos\bigb{\Delta\theta_1} & \sin\bigb{\Delta\theta_1}\\
\vdots & \vdots \\
\cos\bigb{\Delta\theta_N} & \sin\bigb{\Delta\theta_N}
}
\mx{K_1\\K_2}
=
\mx{
\cos\bigb{\theta_1}\\
\vdots\\
\cos\bigb{\theta_N}
}
}
where  \eq{
        \cos\bigb{\theta_i} = \frac{\cos\bigb{\frac{\tilde{x}_i-x_c^*}{r_{min}}}-b^*}{\cos\bigb{\frac{\tilde{x}_i-x_c^*}{r_{min}}}b^*-1}
    }
Lastly, $\theta_0 = \tan^{-1}\bigb{K_2/K_1}$. While this proposed calculation is similar to the fourth step of reconstruction in PLANET \cite{shcherbakova2018planet}, it differs in that $cos(\theta_i)$ is directly calculated.

A flowchart of CELF is given in Fig. \ref{fig:steps}. Implementation of the CELF will be available for general use at http://github.com/icon-lab/mrirecon.

\section{Experiments}

\subsection{Simulations}
To comprehensively assess estimation performance, three separate simulations were performed. First, phase-cycled bSSFP signals were simulated for nine different tissues under varying noise levels. The following tissues were considered: fat, bone marrow, liver, white matter, myocardium, vessels, gray matter, muscle and CSF. The relaxation times of the tissues at 3T were selected according to \cite{bojorquez2017normal}, as listed in Supp. Tab. I. $TR$=$8$ms, $TE$=$4$ms, flip angle $\alpha$=$40^o$ and $N=\{6,8\}$ were assumed. Bi-variate Gaussian noise was added to simulated signals to attain SNR values in the range [20, 100]. SNR values were calculated separately for each tissue, so at a fixed SNR level the standard deviation of noise varied across tissues with respect to the level of tissue signal. SNR was taken as defined in \cite{bjork2014parameter,shcherbakova2018planet}:
\eq{
    \text{SNR} = \frac{\sum_{n=1}^N |S_n|}{N\sigma}
}
where $S_n$ is the signal for the $n^{th}$ phase-cycled acquisition, $\sigma$ is the standard deviation of noise. Monte-Carlo simulations were repeated 10000 times with independent noise instances. At each instance, $\theta_0$ was chosen from uniformly distributed values between $-\pi$ and $\pi$. $T_1$ and $T_2$ estimation was performed via CELF and PLANET. Performance was quantified as mean absolute percentage error (MAPE):
\eq{
    \text{MAPE(\%)} = \frac{100}{K}\sum_{k=1}^{K} \abs*{ \frac{ T_{i,true}^k-T_{i,estimated}^k}{T_{i,true}^k}}
}
where $K$ is the number of repetitions and $T_i^k$ is the value of $T_i$ ($i\in{1,2}$) in the $k^{th}$ repeat.

Second, phase-cycled bSSFP signals were simulated for the same set of tissues for varying flip angles and off-resonance. $TR=8$ms, $TE=4$ms, flip angle $\alpha=20^o-60^o$, off-resonance from -62.5 Hz to 62.5 Hz (i.e., $-0.5/TR$ to $0.5/TR$), and $N=\{4,6,8\}$ were considered. The noise level was adjusted for an SNR of 200 for CSF, and was kept uniform across tissues. Simulations were repeated 20 times with independent noise. $T_1$, $T_2$, off-resonance, and banding-free images were estimated via CELF and PLANET. 

Third, performance of CELF was examined under moderate deviations in flip angle  from its prescribed nominal value. Phase-cycled bSSFP signals were simulated for the same set of tissues under the following parameters: $TR=8$ms, $TE=4$ms, zero additive noise and off-resonant frequency shift, $N=4$. Within each block, the nominal flip angle was varied in [$20^o$ $60^o$] vertically, and the ratio of actual to nominal flip angle was varied in [$0.9$ $1.1$] horizontally (Supp. Fig. 4). CELF was performed on the simulated bSSFP signals, and the percentage error was calculated between the CELF-estimates and true parameter values.

Lastly, the utility of CELF in synthesizing bSSFP images at varying flip angles was assessed. To do this, banding-free bSSFP images at flip angles ($\alpha={20^o, 30^o, 50^o, 60^o}$) were generated based on CELF parameter estimates obtained at $\alpha=40^o$. Phase-cycled bSSFP signals were simulated for the same set of tissue under the following parameters: $TR=8$ms, $TE=4$ms, flip angle, off-resonance varying from -62.5 Hz to 62.5 Hz (i.e., $-0.5/TR$ to $0.5/TR$) horizontally, and $N=\{4,6,8\}$. $T_1$, $T_2$, and banding-free image estimates were obtained via CELF. Banding-free bSSFP images were then generated based on Eq. (\ref{eq:signal_model}). Synthetic images obtained using $N=\{4,6,8\}$ acquisition were compared against reference banding-free bSSFP images directly simulated at the target flip angles ($\alpha={20^o, 30^o, 50^o, 60^o}$).

\begin{figure*}[t]
\centering
\includegraphics[width=\textwidth]{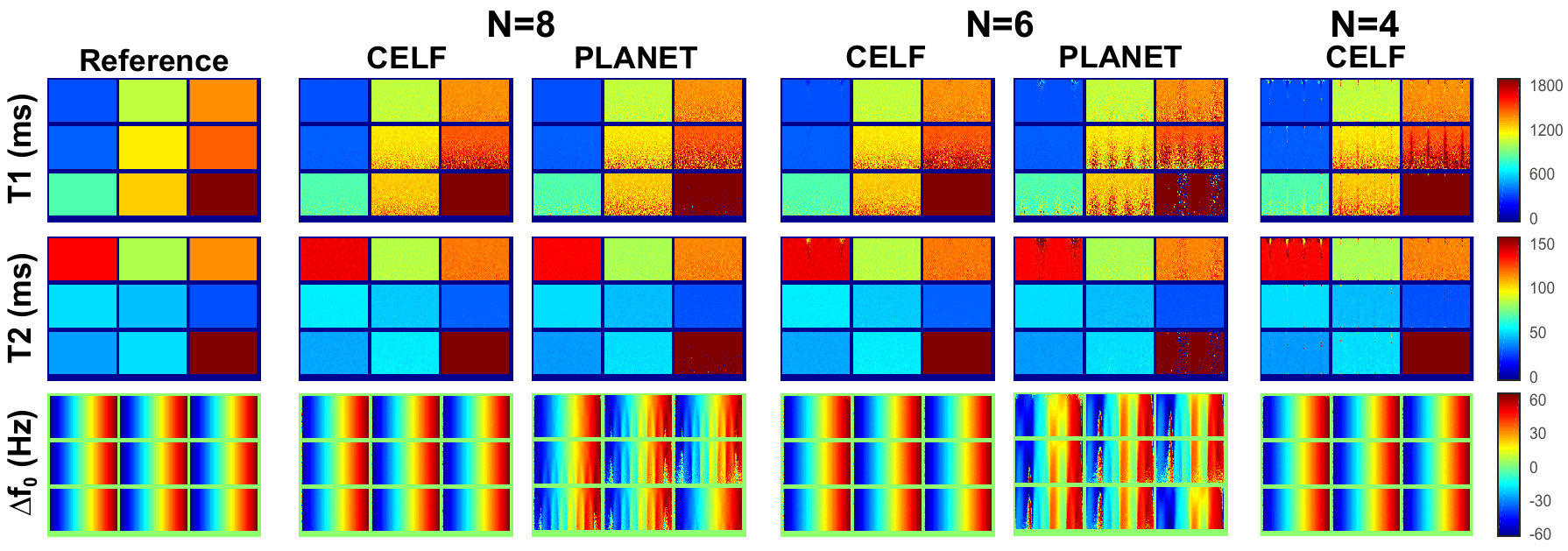} 
\caption{Phase-cycled bSSFP signals were simulated for nine tissue blocks (left column from top to bottom: fat, bone marrow, liver; middle column from top to bottom: white matter, myocardium, vessels; right column from top to bottom: gray matter, muscle, and CSF). In each block off-resonance varied from -62.5 Hz to 62.5 Hz ($1/2.TR$) along the horizontal axis, and flip angle varied from 20$^o$ to 60$^o$ along the vertical axis. Remaining parameters were: $TR=8$ms, $TE=4$ms, $N=\{4,6,8\}$, and SNR = 200 (with respect to CSF signal intensity). $T_1$, $T_2$, and off-resonance estimates via CELF and PLANET are displayed. Note that PLANET cannot compute estimates for N=4.}
\label{fig:sim2} 
\end{figure*}

\subsection{Phantom and In vivo Studies}
Phantom and in vivo experiments were performed on a 3T scanner (Siemens Magnetom Trio, Erlangen, Germany) equipped with gradients of maximum strength 45 mT/m and maximum slew rate 200 T/m/s. Imaging protocols were approved by the local ethics committee at Bilkent University, and informed consent was obtained from all participants. Phase-cycled bSSFP acquisitions were collected without delay using a 3D Cartesian sequence. Standard volumetric shimming was performed at the beginning of the session, and the shim was unaltered thereafter. A subset of acquisitions from an N=8 scan were selected for assessments with N=4 ($\Delta\theta=\{0,\pi/2,\pi,3\pi/2\}$) and N=6 ($\Delta\theta=\{0,\pi/4,\pi/2,\pi,5\pi/4,3\pi/2\}$). Details regarding the scan protocols are listed below.

\textbf{Phantom bSSFP:} Phase-cycled bSSFP acquisitions of a cylindrical phantom were performed using a $12$-channel head coil. The phantom contained a homogeneous mixture of 2.42 mM/L $NiSO_4$ and 8.56 mM/L $NaCl$ in a plastic casing of diameter 115 mm and height 200 mm. The sequence parameters were a $TR/TE$ of $8.89/4.445$ ms, a flip angle of $55^o$, an FOV of $175\text{ mm}\times175\text{ mm}\times120\text{ mm}$, a matrix size of $128\times128\times40$, a readout bandwidth of $160$ Hz/Px, N=8 with $\Delta\theta$ spanning $[0,2\pi)$ in equispaced intervals. Total scan time for N=8 was 6:32. 
    
\textbf{In vivo bSSFP:} Phase-cycled bSSFP acquisitions of the brain were performed using a $32$-channel head coil for two healthy subjects. The sequence parameters were a $TR/TE$ of $8.18/4.09$ ms, a flip angle of $40^o$, an FOV of $256\text{ mm}\times256\text{ mm}\times120\text{ mm}$, a matrix size of $256\times256\times30$, a readout bandwidth of $190$ Hz/Px, N=8 with $\Delta\theta$ spanning $[0,2\pi)$ in equispaced intervals. Total scan time for N=8 was 8:40.

\textbf{Additional scans:} Reference $T_1$ and $T_2$ maps were obtained using gold-standard sequences near the central cross-section of the bSSFP acquisitions. $B_1$ maps were acquired to estimate the actual flip angle across the field-of-view (FOV). The collected $B_1$ maps were denoised using a non-local means filter with the following parameters: $21\times21$ search window, $5\times5$ comparison window, and $0.01$ degree of smoothing. Afterwards, $B_1$ correction was performed to adjust the nominal flip angle at each voxel. $B_0$ maps were acquired to check for potential drifts in the main field inhomogeneity during the session. No significant main field drift was observed during the scans, so $B_0$ maps were not utilized for correction. Please see Supp. Text 2 for further details. 

An adaptive coil-combination was performed on multi-coil images to produce a complex-valued image \cite{walsh2000adaptive}. All analyses were performed voxel-wise (with the exception of singular voxels) on this coil-combined image using MATLAB R2015b (The MathWorks, Inc., MA, USA). The adaptive combination performs local block-wise combinations based on spatial-matched filters estimated from signal and noise correlation matrices. For a given block, the signal correlation matrix is estimated on a broader surrounding neighborhood, while the noise correlation matrix is estimated on peripheral regions without tissues. The stock implementation of the adaptive-combination method was adopted with default parameters on the 3T Siemens platform used here. $T_1$, $T_2$, off-resonance, and banding-free images were estimated via CELF and PLANET. A variant of PLANET (PLANET+$\Delta f_0$) was also implemented to adopt the off-resonance step from CELF. Tissue-specific evaluation of parameter estimates was performed on white matter, gray matter and CSF region-of-interests (ROI) manually defined in each subject (see Supp. Figs. 5 and 6). ROIs were selected as spatially-contiguous regions containing homogeneous parameter values.

To demonstrate CELF's utility in synthesizing bSSFP images, banding-free bSSFP images at flip angles ($\alpha={20^o, 30^o, 50^o, 60^o}$) were generated based on CELF parameter estimates obtained at $\alpha=40^o$. Separate sets of $T_1$, $T_2$, and banding-free image estimates were obtained via CELF at $N=\{4,6,8\}$. Banding-free bSSFP images at target flip angles were then generated based on Eq. (\ref{eq:signal_model}).

Finally, we explored the feasability of a learning-based correction for CELF parameter estimates. Artificial neural networks were recently demonstrated to reproduce non-bSSFP relaxometry maps from phase-cycled bSSFP signals \cite{heule2020multi}. Inspired by this recent method, we reasoned that neural network models can be trained to predict reference parameter maps given CELF-derived parameter estimates as input, and that the trained models can further improve accuracy as a post-correction step to CELF. For proof-of-concept demonstrations, data from four subjects were analyzed. Model training was performed on three subjects, and the resulting model was tested on the held-out subject; and this procedure was repeated for all possible training sets. A fully-connected feedforward architecture was used with a total of 942 parameters. This architecture comprised an input layer that received CELF estimates for $T_{(1,2)}$, 3 hidden layers of 20 neurons each and sigmoid activations, an output layer with 2 neurons and linear activations for final $T_{(1,2)}$ predictions. The network performed voxel-wise processing, where CELF parameter estimates at N=8 were taken as input and parameter estimates from the gold-standard mapping sequence were taken as ground truth. $T_1$ and $T_2$ values were independently normalized to a maximum of 1 across voxels in the training set, and this global scaling factor was reapplied to the network output during testing. Model training was performed using the RProp algorithm to minimize mean absolute error between network predictions and ground truth \cite{Riedmiller1993}. Network weights and bias terms were randomly initialized as uniform variables in the range [-1 1]. Training data were split randomly, with 15\% of voxels held-out as validation set. Network training was stopped when the error in the validation set did not diminish in 10 consecutive epochs.

\section{Results}

\begin{figure*}[t]
\centering
\includegraphics[width=\textwidth]{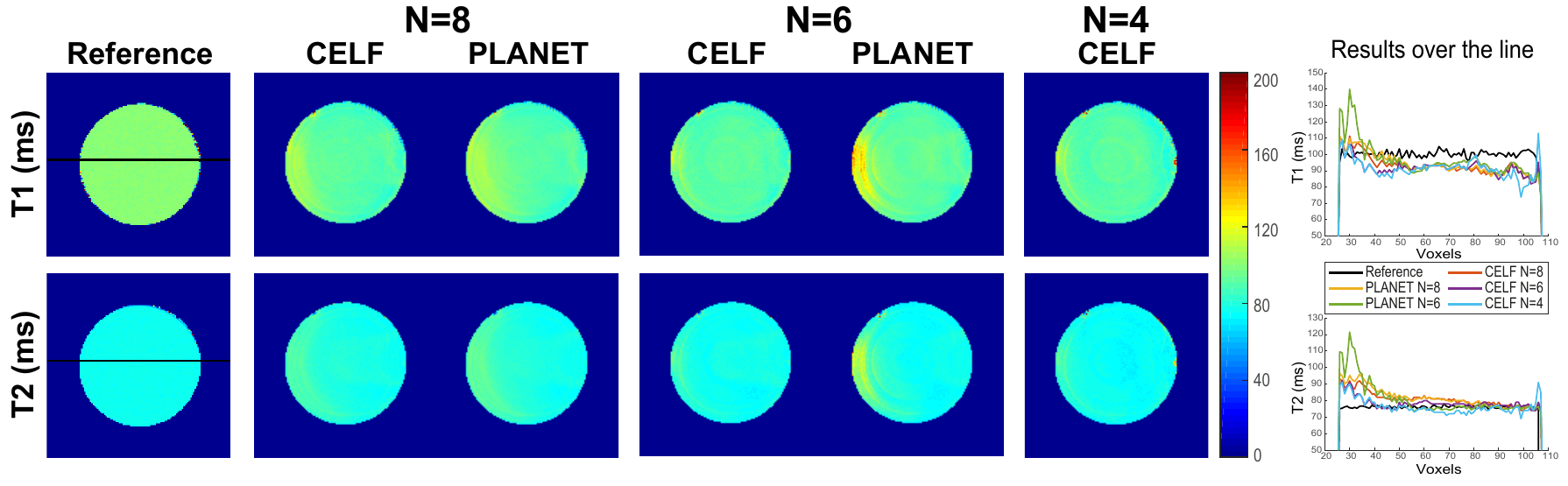}
\caption{\label{fig:phantom_res} $T_1$ and $T_2$ estimation was performed with CELF and PLANET on phase-cycled bSSFP images of a homogeneous phantom. Results are shown in a representative cross-section for $N={\{4,6,8\}}$. Note that PLANET cannot compute estimates for $N={\{4\}}$. $T_1$ and $T_2$ estimates are displayed along with the reference maps from gold-standard mapping sequences. To better visualize differences among methods, variation of estimates are also plotted across a central line (black line in reference maps).}
\end{figure*}

\subsection{Simulations}
The proposed method was first evaluated on simulated bSSFP signals. Parameter estimation was performed for varying SNR levels ([20 100]) and number of acquisitions ($N={\{4,6,8\}}$ for CELF, $N={\{6,8\}}$ for PLANET). Mean absolute percentage of estimation error (MAPE) was calculated separately for each SNR, tissue type and N. Supp. Fig. 7 displays MAPE for $T_1$ and $T_2$ as a function of SNR. For all tissue types, CELF yields lower MAPE compared to PLANET consistently across SNR levels. Most substantial differences are observed in $T_1$ estimation for tissues with relatively higher $T_1/T_2$ ratio, including liver, myocardium, muscle and vessels. In general, the performance difference between the two methods diminishes towards higher SNR, and tissues with similar $T_1/T_2$ ratio exhibit similar errors.

Next, evaluations were performed for $T_1$, $T_2$, off-resonance, and banding free image estimation on a numerical phantom. The numerical phantom comprised nine tissue blocks, off-resonance varied in $\theta_0 \in [ -\pi, \pi ]$ horizontally and flip angle varied in [20 60]$^o$ vertically within each block (see Supp. Fig. 8 for phase-cycled bSSFP images). Estimates were obtained with CELF ($N={\{4,6,8\}}$) and PLANET ($N={\{6,8\}}$). Fig. \ref{fig:sim2} shows the estimation results for $T_1$, $T_2$, off-resonance terms for each N along with reference ground-truth parameter maps (see Supp. Fig. 9 for estimates of banding-free images, Supp. Fig. 10 for comparison of results with and without singularity detection). CELF produces more accurate maps than PLANET for $T_1$, $T_2$, and off-resonance terms. Performance benefits with CELF become more noticeable towards lower N, and in off-resonance estimation. 

To examine effect of $B_1$ inhomogeneity on CELF estimates, a separate set of $T_1$, $T_2$, off-resonance, and banding-free image estimates was obtained in the numerical phantom. In this case, the nominal flip angle varied in [20 60]$^o$ vertically, and the ratio of actual to nominal flip angle varied in [0.9 1.1] horizontally within each block. Percentage error in parameter estimates is displayed in Supp. Fig. 11. Averaged across tissue blocks, the mean estimation errors are $\%11.11$ for $T_1$, $\%0.0$ for $T_2$, $\%2.32$ for banding-free image, and $\%0.0$ for off-resonance estimates. We observe that $T_2$ and off-resonance estimates are not affected. Note that this is in contrast to PLANET that shows elevated errors in off-resonance estimates due to flip angle deviations (Supp. Fig. 12). Meanwhile, banding-free images and $T_1$ estimates show modest effects due to disparity between nominal and actual flip angles. Regardless, the proposed procedure for CELF includes a $B_1$-mapping step to surmount potential disparities. 

Lastly, we demonstrated CELF's utility in synthesizing bSSFP images at varying flip angles. Phase-cycled bSSFP images of the numerical phantom were simulated for matching nominal and actual flip angles of $\alpha=40^o$. $T_1$, $T_2$, and banding free image estimates at flip angle $=40^o$ were obtained with CELF. CELF parameter estimates were then used to synthesize banding-free bSSFP images at different flip angles $\{20^o, 30^o, 50^o, 60^o\}$. The CELF-derived bSSFP images at $N=\{4,6,8\}$ were compared against reference images directly simulated at the respective flip angles (Supp. Fig. 13). CELF-derived synthetic bSSFP images are virtually identical to reference bSSFP images.

\begin{table}[b]
\caption{Estimated $T_1$ and $T_2$ values for in vivo brain images \label{tab:invivo_roi}}
\centering
\resizebox{\columnwidth}{!}{ 
\begin{tabular}{|c|c|c|c|c|c|c|c|} 
\hline 
\multicolumn{2}{|c|}{\multirow{2}{*}{\textbf{Methods}}} & \multicolumn{2}{c|}{\textbf{WM}} & \multicolumn{2}{c|}{\textbf{GM}} & \multicolumn{2}{c|}{\textbf{CSF}} \\ \cline{3-8} 
\multicolumn{2}{|c|}{} & $T_1$ (ms) & $T_2$ (ms) & $T_1$ (ms) & $T_2$ (ms) & $T_1$ (ms) & $T_2$ (ms) \\ \hline
\multirow{3}{*}{\rotatebox[origin=c]{0}{CELF}} & N=8 & 411 $\pm$ 55 & 50 $\pm$ 8 & 849 $\pm$ 239 & 62 $\pm$ 25 & 2685 $\pm$ 1068 & 1298 $\pm$ 660 \\\cline{2-8}
& N=6 & 462 $\pm$ 63 & 50 $\pm$ 7 & 800 $\pm$ 166 & 63 $\pm$ 16 & 2828 $\pm$ 1011 & 1313 $\pm$ 601 \\\cline{2-8}
& N=4 & 416 $\pm$ 65 & 51 $\pm$ 8 & 795 $\pm$ 141 & 63 $\pm$ 16 & 3485 $\pm$ 398 & 1665 $\pm$ 536 \\\hline
\multirow{2}{*}{\rotatebox[origin=c]{0}{PLANET}} & N=8 & 396 $\pm$ 67 & 48 $\pm$ 8 & 574 $\pm$ 2255 & 37 $\pm$ 227 & - & - \\\cline{2-8}
& N=6 & 384 $\pm$ 72 & 47 $\pm$ 8 & 763 $\pm$ 466 & 60 $\pm$ 22 & - & - \\\hline
\multicolumn{2}{|c|}{Reference} & 622 $\pm$ 23 & 63 $\pm$ 4 & 870 $\pm$ 70 & 56 $\pm$ 5 & 3412 $\pm$ 1148 & 613 $\pm$ 340 \\\hline 
\end{tabular}
}
\end{table}

\subsection{Phantom and In vivo Studies}
To validate the proposed method, phase-cycled bSSFP acquisitions were performed on a homogeneous phantom. Phantom images for individual phase cycles are shown in Supp. Fig. 14. $T_1$ and $T_2$ maps were then estimated with  CELF ($N={\{4,6,8\}}$) and PLANET ($N={\{6,8\}}$). Estimates in a representative cross-section are presented in Fig. \ref{fig:phantom_res} along with reference maps obtained via conventional mapping sequences. Compared to PLANET, CELF produces more accurate results with improved homogeneity of parameter estimates across the phantom. Moreover, parameter estimates by CELF are highly consistent across $N={\{4,6,8\}}$ as listed in Supp. Table III. 

\begin{figure*}[t]
\centering
\includegraphics[width=\textwidth]{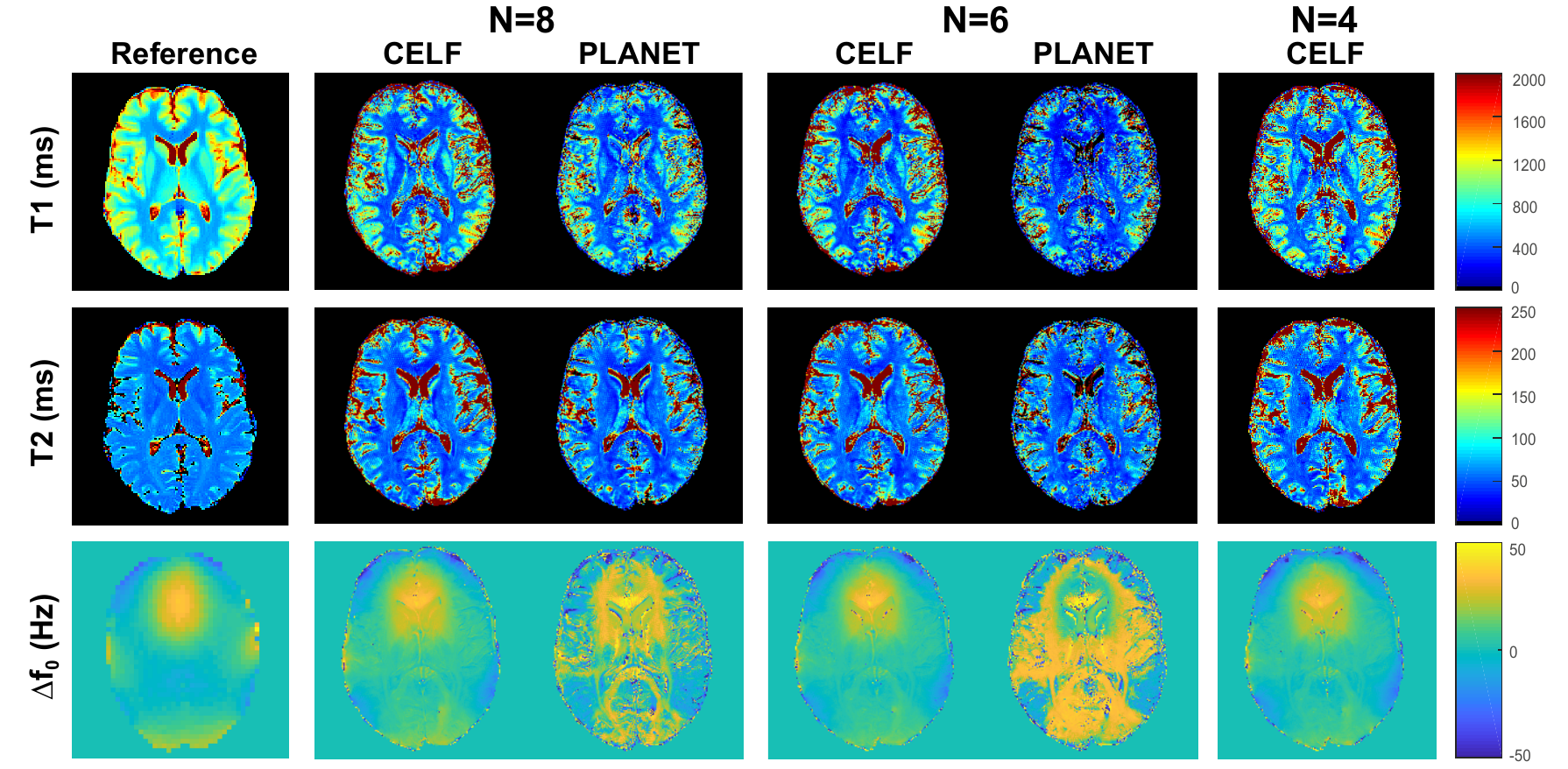}
\caption{\label{fig:invivo_res} Parameter estimation was performed with CELF and PLANET on phase-cycled bSSFP images of the brain. Results are shown in a cross-section of a representative subject for $N={\{4,6,8\}}$. Note that PLANET cannot compute estimates for $N={\{4\}}$. $T_1$, $T_2$, and off-resonance estimates are displayed along with the reference maps from gold-standard mapping sequences.}
\end{figure*}

Next, in vivo phase-cycled bSSFP brain images were acquired in two subjects (see Supp. Figs. 15 and 16 for individual phase-cycles). $T_1$, $T_2$ and off-resonance maps were estimated along with banding-free bSSFP images. Estimations were performed with CELF ($N={\{4,6,8\}}$) and PLANET ($N={\{6,8\}}$). Estimated maps in representative axial cross-sections are displayed in Fig. \ref{fig:invivo_res} and Supp. Fig. 17 for Subject 1, and in Supp. Fig. 18 for Subject 2. (See Supp. Fig. 19 for location of bounded search procedure for $\gamma$ in Subjects 1 and 2.) For $T_1$ and $T_2$ estimation, CELF-derived maps show visually lower levels of noise and residual artifacts than than PLANET-derived maps, particularly near tissue boundaries. This improvement can be partly attributed to the dictionary-based ellipse identification in CELF as illustrated in Supp. Fig. 20. For off-resonance estimation, CELF yields noticeably more accurate maps than PLANET, and this difference grows towards lower $N$. This improvement can be attributed to the improved procedure in CELF for $B_o$ estimation as demonstrated in Supp. Figs. 21 and 22.

To assess the synthesis ability of CELF, banding-free bSSFP images at varying flip angles were generated based on CELF-parameter estimates obtained for a fixed flip angle. Specifically, $T_1$, $T_2$, and banding-free image estimates at flip angle $=40^o$ were obtained with CELF. CELF parameter estimates were then used to synthesize banding-free bSSFP images at different flip angles $\{20^o, 30^o, 50^o, 60^o\}$ (Supp. Figs. 23 and 24). CELF-derived synthetic bSSFP images are virtually identical to reference bSSFP images. As theoretically expected, CSF appears brighter towards higher flip angles resulting in bright fluid contrast, whereas gray-matter and white-matter signals diminish in intensity.

Average $T_1$ and $T_2$ values across white matter (WM), grey matter (GM), and CSF ROIs in a central cross-section are listed in Table \ref{tab:invivo_roi}, Supp. Tables IV-VI for Subjects 1-4. PLANET yielded negative parameter estimates for many CSF voxels, so no measurements were reported. Overall, CELF and PLANET yield consistent results in WM and GM ROIs. There is also good agreement between CELF-estimated and reference $T_2$ values for WM and GM. Meanwhile, $T_1$ is moderately underestimated in WM and GM. Note that although CELF-derived $T_2$ values are higher than reference $T_2$ values in CSF, CELF-derived values are in better accordance with prior studies reporting $T_2$ values at least greater than 1500 ms \cite{spijkerman2018t}.

Lastly, we explored the feasibility of learning-based post-correction to CELF for enhanced performance in relaxation parameter estimation. A neural-network model was trained to receive as input CELF-derived $T_1$ and $T_2$ estimates, and output reference parameter values from the gold-standard mapping sequences. Cross-validated parameter estimates were obtained in each subject. Representative parameter maps following the correction are displayed in Supp. Fig. 25. Percentage difference between the CELF-based parameter estimates and reference parameter values before and after the correction are listed in Supp. Table VII. Across subjects, the average discrepancy between CELF-based and reference parameter values is reduced from 22.8\% to 7.9\% in WM, from 13.4\% to 11.8\% in GM, and from 46.7\% to 34.5\% in CSF. These results imply that a model-based learning method that combines CELF with neural networks can offer high estimation accuracy in WM and GM particularly.

\section{Discussion}

In this study, we introduced a constrained ellipse fitting (CELF) method for parameter estimation in phase-cycled bSSFP imaging. CELF devises an ellipse fit with a reduced number of unknowns by leveraging prior knowledge on the geometrical properties of the bSSFP signal ellipse. Specifically, the central line and orientation of the ellipse are considered to enable ellipse fits with a minimum of 4 (see Eq.~\eqref{eq:costQ}) as opposed to 6 (see Eq.~\eqref{eq:defCost}) data points. The proposed method was demonstrated for estimation of $T_1$, $T_2$ and off-resonance values as well as banding-free bSSFP images from phase-cycled bSSFP acquisitions. CELF shows improved estimation accuracy compared to PLANET. It also offers higher scan efficiency by reducing the number of phase-cycles required, albeit it employs a separate $B_1$ map to mitigate sensitivity to flip angle variations. We find that CELF yields accurate estimates of off-resonance and banding-free bSSFP images with as few as N=4 acquisitions. At the same time, in vivo estimates of relaxation parameters with CELF show biases compared to conventional spin-echo methods, due to microstructural sensitivity of bSSFP imaging. Further development of CELF is warranted to resolve these estimation errors, which might eventually provide useful opportunities for CELF in quantitative MRI applications including relaxometry, magnetization transfer (MT) mapping \cite{wood2020magnetization} and conductivity mapping \cite{gavazzi2020transceive}.

\subsection{Parameter Estimates}
Analyses performed on simulated bSSFP signals indicate that both PLANET and CELF yield higher performance in $T_1$ and $T_2$ estimation at relatively low flip angles. This property can facilitate bSSFP parametric mapping by reducing specific absorption rate (SAR), particularly at higher field strengths. Performance losses are observed towards relatively high flip angles, where the bSSFP signal intensities for many tissues diminish. Furthermore, the ellipse shape tends towards a circle for higher flip angles, diminishing shape differences among different tissues and increasing difficulty of $T_1$, $T_2$ estimation. While PLANET yields relatively improved off-resonance estimates towards low flip angles, CELF maintains consistent performance across a broad range of flip angles and N. This difference is also observed for in vivo experiments, clearly demonstrating the benefit of the off-resonance estimation step in CELF. Reducing error propagation in parameter estimation by removing intermediate steps renders the off-resonance estimate of CELF more robust to variability in N and flip angle. This robust estimation of off-resonance might be helpful in transceive phase mapping \cite{gavazzi2020transceive}.

Here, CELF accurately estimated off-resonance maps and produced banding-free bSSFP images in all scenarios considered. Moderate aggreement was also observed between estimated and reference relaxation parameters for simulation and phantom experiments. One exception was the parameter estimates for the homogeneous cylindrical phantom, where artefactually high parameter estimates were observed towards the left rim. It is likely that the estimation errors are induced by inhomogeneous intravoxel frequency distribution due to mechanical vibrations and residual air within the phantom. There was also some apparent discrepancy for in vivo $T_1$-$T_2$ estimates from both PLANET and CELF, particularly for white matter. Our observations match well to previous studies on bSSFP-based relaxometry reporting similar effects in the human brain \cite{nguyen2017motion,shcherbakova2018planet}. This discrepancy is best attributed to the intravoxel frequency distribution when multiple tissue compartments reside in a single voxel. Frequency inhomogeneities in multi-compartment voxels in turn induce asymmetries in the bSSFP signal profile \cite{miller2010asymmetries,miller2010asymmetries2}. Such asymmetries are not evident in bSSFP signals during simulation or phantom experiments, but bSSFP signals in the brain reflect contributions from both micro- and macro-structural components such as myelin and water. Therefore, the single compartment bSSFP signal model used in CELF can be affected by the microstructural sensitivity of bSSFP imaging.

CELF based on a single-compartment model with inherent microstructural sensitivity can have several practical applications. The microstructural bias on bSSFP signals is expected to be relatively limited at lower field strenghts \cite{miller2010asymmetries}. As such, one application domain for CELF is parameter mapping in the brain using low-field systems, particularly 1.5 T or lower. While relaxation parameter estimates can carry some microstructural bias, off-resonance maps derived by CELF show strong correlation to reference maps. Therefore, the enhanced reliability of off-resonance estimates with CELF can improve applications that rely on assessment of signal phase such as electrical impedance tomography. CELF can also be leveraged to produce synthetic bSSFP images at varying flip angles, to improve tissue delineation at multiple distinct contrasts \cite{hilbert2018true}. Lastly, the intrinsic microstructural sensitivity in CELF can be viewed as an additional contrast mechanism, and concurrent analysis of spin-echo and bSSFP-based parameter maps can allow estimation of tissue microstructural properties \cite{heule2020multi}.

Several approaches can be adopted to alleviate the microstructural bias in CELF estimates when necessary. A two-compartment model was recently proposed for PLANET to improve accuracy for parameter estimation \cite{shcherbakova2019accuracy}. Similarly, the base bSSFP signal equation in CELF can be generalized via a multi-compartment model to enhance parameter estimates in the brain, albeit a greater number of measurements might be necessary due to increased degrees of freedom in the model. Alternatively, learning-based methods can be leveraged in conjunction with CELF to reduce discrepancy in relaxation parameter estimates. Here, a proof-of-concept demonstration was provided suggesting the potential for enhanced estimation accuracy with neural-network-based post-correction of CELF estimates. Demonstrations on broader datasets are needed to fully outline the potential benefits of this model-based learning approach. 

\subsection{Potential Improvements and Future Work}
CELF performs constrained ellipse fits based on an analytical model of the bSSFP signal, followed by parameter estimation via dictionary-based ellipse identification. Similar to the PLANET formulation, the analytical model used for the fits and simulation of the ellipse dictionary assume negligible non-stationary effects on measurements from motion, eddy currents or main field drifts. In theory, incorporating additional phase-cycled measurements during ellipse fitting should improve estimation accuracy for CELF. Yet, CELF estimates for relaxation parameters were occasionally observed to be somewhat more accurate at lower N. Note that bSSFP protocols with greater N increase the risk of differential measurement biases among phase cycles due to non-stationary effects. Such perturbations could then counter the benefits of additional measurement points on ellipse fits. In cases where substantial biases are present, biasing factors can be separately measured or else estimated from bSSFP data, and the bSSFP measurements can be corrected a priori to improve correspondence with the underlying signal model. Note that CELF allows lower N that inherently reduces potential for motion artifacts. That said, parameter estimates can be suboptimal under severe motion. A potential solution is to spatially align images for separate phase-cycles prior to ellipse fitting. Sensitivity to subject motion can also be limited by accelerating individual bSSFP acquisitions \cite{ccukur2015accelerated,ilicak2017profile,biyik2018reconstruction}. To check for drifts in the main field, $B_0$ maps were collected at the beginning and end of the scan sessions. No significant effects were observed, so no correction procedures were applied here. Yet main field changes might be an important factor for high-resolution and large-FOV protocols that require prolonged scan times. In such cases, the recently proposed $B_0$ drift-correction method can be applied to improve estimation accuracy \cite{shcherbakova2019investigation}.

The current CELF implementation assumes that noise is independent and identically distributed across separate phase-cycled acquisitions. If the noise distribution varies substantially across acquisitions, the loss terms in the ellipse fitting or identification stages might be biased by higher levels of noise in a subset of measurements. Note that both ellipse fitting errors between the analytical signal model and measurements, and ellipse identification errors between the dictionary ellipse and measured ellipse are accumulated across points on the ellipse (i.e., individual phase-cycled acquisitions). Therefore, to minimize bias due to non-stationary noise, the loss terms can be modified to weigh errors from individual points inversely with their noise level.

Here parameter estimation was performed separately for each voxel in phase-cycled bSSFP images, omitting potential correlations among neighboring voxels. In the presence of high measurement noise, voxel-wise processing might lead to spatially unsmooth estimates even in regions with a homogeneous distribution of parameters. In such cases, estimation accuracy can be improved by recasting the optimization problem in CELF to enforce locally-coherent parameter estimates at the expense of potential blurring. Alternatively, non-local means (NLM) filtering can be applied on the CELF-derived parameter maps to improve accuracy without excessive smoothing \cite{manjon2010adaptive}. Note that spatially distant tissues with similar $T_1$ and $T_2$ values could still manifest different off-resonance values. Since CELF back-rotates ellipses to remove off-resonance effects prior to ellipse fitting, it is well suited to NLM methods.

Simulation results suggest that CELF incurs moderate performance losses in $T_1$ estimation with up to \%20 mismatch between the nominal and actual flip angles. Given this sensitivity to flip angles, accuracy of $T_1$ estimates rely on knowledge of the spatial distribution of the flip angle. To mitigate potential errors, in this study, we obtained $B_1$ maps to correct the nominal flip angle at each voxel. Since additional $B_1$ mapping scans lower scan efficiency, they may not be desirable in all protocols. Moreover, accurate $B_1$ mapping might be relatively difficult in higher field systems. In those cases, generalizing CELF to additionally estimate flip angle maps would improve estimation accuracy.

In this study, both PLANET and CELF implementations performed parameter estimation on coil-combined complex bSSFP images. To do this, multi-coil bSSFP images were combined using the adaptive combination method prior to ellipse fitting. This method can offer better noise suppression than the conventional sum-of-squares method in regions of lower signal intensity, and it reduces computational burden by lowering data dimensionality across the coil dimension. That said, prior studies suggest that coil combination methods primarily devised for magnitude images might introduce instabilities in reconstructed phase images \cite{Robinson2011}. Thus, accuracy of ellipse fits based on coil-combined images from the adaptive combination method might be limited by instability-driven phase variability across phase-cycles. It remains important future work to inquire the potential benefits of more advanced approaches for processing of multi-coil information in conjunction with the elliptical signal models \cite{McKibben2019,Xiang2020}.

Compared to PLANET that involves direct ellipse fitting, CELF has an added dictionary-based ellipse identification step to refine the estimates. As a result, CELF run times also reflect the computational cost of this identification step. Note that the ellipse dictionary can be constructed once a priori, and so the inference cost stems from the search process to match the bSSFP signal ellipse to the closest dictionary ellipse. Here a brute-force search was performed since parameter maps could be extracted within several minutes for all datasets considered. When desired, the computational efficiency of CELF can be boosted by leveraging advanced matching procedures \cite{Cauley2015} or GPU accelerated implementations \cite{Wang2020}.

In summary, we introduced the analytical foundation of a constrained ellipse fitting procedure (CELF) for bSSFP-based parameter estimation. Given phase-cycled bSSFP images along with a separate $B_1$ map, CELF can estimate relaxation parameters, off-resonance and banding-free bSSFP images. Here, we demonstrated CELF using simulated, phantom and in vivo experiments. Our results indicate that CELF achieves improved efficiency compared to direct ellipse fitting. CELF can produce accurate estimates of off-resonance and banding-free bSSFP images with as few as N=4 acquisitions, yet in vivo estimation accuracy of relaxation parameters is limited due to microstructural sensitivity of bSSFP imaging. Further work to mitigate errors in relaxation parameter estimates is needed to render CELF a candidate for relaxometry applications. Broader assessments on physical phantoms with systematic variation of field inhomogeneity and relaxation parameters, and on patients with pathology are critical steps towards validation. CELF might contribute to the clinical utility of bSSFP imaging following these technical improvements and validations, as it can improve scan efficiency and reliability against main field inhomogeneity.

\newpage
\printbibliography[heading=subbibliography] 
\end{refsection}




\clearpage

\renewcommand{\shorttitle}{}

\newcommand{\T}{\rule{0pt}{2.6ex}}
\newcommand{\B}{\rule[-1.0ex]{0pt}{0pt}}
\renewcommand{\thetable}{\Roman{table}}

\newif\ifsupp
\supptrue

\linespread{1.5}
\newcommand{\ohwidth}{0.75\columnwidth}
\newcommand{\swidth}{0.5\columnwidth}
\newcommand{\dwidth}{\columnwidth}
\newcommand{\mwidth}{0.6\columnwidth}

\setcounter{tocdepth}{4}
\setcounter{secnumdepth}{4}

\let\citeleft=(
\let\citeright=)

\crefname{equation}{eq.}{eqs.}
\crefname{figure}{fig.}{figs.}
\crefname{table}{table}{tables}

\setlength{\parskip}{4mm plus1mm minus1mm}

\begin{center}
	\begin{Large}
		\begin{bf}
Supporting Document For:\\[1em]
Constrained Ellipse Fitting for Efficient Parameter Mapping with Phase-cycled bSSFP MRI
		\end{bf}
	\end{Large}
\end{center}
\bigskip
\begin{center}
Kübra Keskin$^{1}$, Uğur Yılmaz$^{2}$, Tolga \c{C}ukur$^{2,3,4}$
\end{center}
\vspace*{0.1in}
\noindent
$^1$Department of Electrical and Computer Engineering, University of Southern California, CA, USA\\
$^2$National Magnetic Resonance Research Center (UMRAM), Bilkent University, Ankara, Turkey\\
$^3$Department of Electrical and Electronics Engineering, Bilkent University, Ankara, Turkey\\
$^4$Neuroscience Program, Sabuncu Brain Research Center, Bilkent University, Ankara, Turkey

\vspace*{0.1in}
\noindent
                                                           
\noindent
{\em Address correspondence to:} \\
	Tolga \c{C}ukur \\
	Department of Electrical and Electronics Engineering, Room 304 \\
	Bilkent University \\
	Ankara, TR-06800, Turkey \\
	TEL: +90 (312) 290-1164 \\
    E-MAIL: cukur@ee.bilkent.edu.tr

\vspace*{0.2in}

\noindent

\vspace*{0.2in}
\noindent
Supporting Table Count:  7 \\
Supporting Figure Count: 25\\

\setcounter{equation}{0}
\setcounter{figure}{0}
\setcounter{table}{0}
\setcounter{page}{1}
\setcounter{section}{0}
\makeatletter
\renewcommand{\theequation}{S. \arabic{equation}}
\renewcommand{\thefigure}{\arabic{figure}}
\captionsetup[table]{name=Supporting Table}
\captionsetup[figure]{name=Supporting Figure}

\clearpage
\begin{refsection}
\section{Supp. Text 1}

\subsection{Subproblem 1: Derivation of $h_{opt}$} \label{appx:step1}
The first subproblem in Eq. (19) can be expressed as:
\eqal{\label{eq:app1}
    \mino{h} \;& Q_(\gamma,\bm{u},h) \\
    \equiv \mino{h} &\; 
    \bnorm{\bigb{\mx{D_0 & \bm{1}_N} - \gamma \mx{D_1 & \bm{0}_N} } \mx{\bm{u} \\ h} }^2_2 \\
    \equiv \mino{h} &\;
    \bigb{(D_0 - \gamma D_1) \bm{u} + \bm{1}_N h}^T\bigb{(D_0 - \gamma D_1) \bm{u} + \bm{1}_N h} \\ 
    \equiv \mino{h} &\; \bm{u}^T(D_0 - \gamma D_1)^T(D_0 - \gamma D_1)\bm{u} + 2\bm{u}^T(D_0 - \gamma D_1)^T\bm{1}_N h + h^2N 
}

The above minimization can be performed by setting the derivative of the cost function with respect to $h$ to zero:
\eqal{
    2\bm{u}^T(D_0 - \gamma D_1)^T\bm{1}_N + 2hN = 0 \\
}
This equation then yields the analytical solution of $h_{opt}$:
\eqal{ h_{opt}=-\frac{1}{N} \bm{1}_N^T(D_0-\gamma D_1)\bm{u} }

\subsection{Subproblem 2: Derivation of $Q_2$} \label{appx:step2_deriveQ2}
By substituting $h_{opt}$ to $Q$ in Eq.~\eqref{eq:app1}, the cost function $Q_2$ can be obtained:
\eqal{
    Q(\gamma,\bm{u},h_{opt}) &= \bm{u}^T(D_0 - \gamma D_1)^T(D_0 - \gamma D_1)\bm{u} + 2\bm{u}^T(D_0 - \gamma D_1)^T\bm{1}_N h + h^2N \\
    &= \bm{u}^T(D_0 - \gamma D_1)^T(D_0 - \gamma D_1)\bm{u}  -\frac{2}{N} \bm{u}^T(D_0 - \gamma D_1)^T\bm{1}_N \bm{1}_N^T(D_0-\gamma D_1)\bm{u} + \frac{1}{N} \bigb{\bm{1}_N^T(D_0-\gamma D_1)\bm{u}}^2 \\
    &= \bm{u}^T(D_0 - \gamma D_1)^T(D_0 - \gamma D_1)\bm{u}  -\frac{1}{N} \bm{u}^T(D_0 - \gamma D_1)^T\bm{1}_N \bm{1}_N^T(D_0-\gamma D_1)\bm{u} \\
    &= \bm{u}^T(D_0 - \gamma D_1)^T(I_N-\frac{1}{N}\bm{1}_N \bm{1}_N^T)(D_0 - \gamma D_1)\bm{u}\\
    &= \bm{u}^T(D_0 - \gamma D_1)^T Z (D_0 - \gamma D_1)\bm{u}\\
    &= \bm{u}^T \bigb{D_0^T Z D_0 - \gamma D_0^T Z D_1 - \gamma D_1^T Z D_0 + \gamma^2 D_1^T Z D_1} \bm{u}\\
    &= \bm{u}^T \bigb{C_0 + \gamma C_1 + \gamma^2 C_2} \bm{u}\\
    &= Q_2(\gamma,\bm{u})
}

\subsection{Subproblem 2: Derivation of the GEP and its solution} \label{appx:step2_GEP}
The constrained optimization problem in Eq. (22) has the following Lagrangian form:
\eqal{
\mathcal{L}(\bm{u},\lambda;\gamma) &= \bm{u}^T \bigb{C_0 + \gamma C_1 + \gamma^2 C_2} \bm{u} - \lambda(\bm{u}^TB\bm{u}-1) \\
}
where $\lambda \in \mathbb{R}$ is the Lagrange multiplier. Note that we initially exclude the strict inequality constraint from the solution set to enforce KKT conditions, but we will later find a feasible solution that satisfies the strict inequality. 

The optimization can be performed by taking the derivative of the Lagrangian and setting it to zero:
\eqal{
\frac{\partial \mathcal{L}}{\partial \bm{u}} = 2C_0\bm{u} + 2\gamma C_1\bm{u} + 2\gamma^2 C_2\bm{u} - 2\lambda B\bm{u} &= \bm{0}_2 \\
\bigb{C_0 + \gamma C_1\ + \gamma^2 C_2 - \lambda B}\bm{u} &= \bm{0}_2
}
Note that the above equation corresponds to a generalized eigenvalue problem (GEP):
\eqal{
\bigb{C_0 + \gamma C_1 + \gamma^2 C_2} \bm{u} = \lambda B \bm{u}
}
where $\lambda$ denotes the eigenvalue and $\bm{u}$ denotes the eigenvector.

To enforce constraints under Karush-Kuhn-Tucker (KKT) conditions we first multiply both sides of Eq. (23) from the left with $\bm{u}_{0}^T$, where $\bm{u}_{0}$ is taken to be a feasible solution: 
\eqal{
\underbrace{\bm{u}_{0}^T \bigb{C_0 + \gamma C_1 + \gamma^2 C_2} \bm{u}_{0}}_{=Q_2(\gamma,\bm{u}_0)} &= \lambda \underbrace{\bm{u}_{0}^T B \bm{u}_{0}}_{=1}
}
Because $Q_2(\gamma,\bm{u}_{0})\geq0$, only non-negative $\lambda$ values are feasible. Furthermore, $G(\gamma) = C_0 + \gamma C_1 + \gamma^2 C_2$ is positive definite or positive semi-definite. These two scenarios are treated separately:
\begin{itemize}
    \item If $G(\gamma)$ is positive definite: \\
    Since eigenvalues of the matrix $B$ are $(-2,2)$, eigenvalues of the GEP will include one negative and one positive value by Sylvester's Law of Inertia \cite{sylvester1852xix}. Therefore, the positive eigenvalue that is also the maximum eigenvalue $\lambda_{max}$ will be the solution.
    \item If $G(\gamma)$ is positive semi-definite: \\
    As at least one eigenvalue should be zero for positive semi-definiteness, there are three configurations for the signs of the eigenvalues: $(\lambda_1,\lambda_2)=(-,0),(0,0),(0,+)$. For $(-,0)$, $0$ will be the solution (the maximum eigenvalue $\lambda_{max}$). For $(0,0)$, no solutions exists since $Q_2(\gamma,\bm{u}_0)=0$ for all $\gamma$. For $(0,+)$, the positive eigenvalue will be the solution.
\end{itemize}
In all scenarios, the largest eigenvalue $\lambda_{max}$ gives a feasible solution under KKT, such that $$Q_3(\gamma) = \lambda_{max}\bigb{C_0 + \gamma C_1 + \gamma^2 C_2,B}$$.

To find a feasible solution that strictly satisfies the constraints $\bm{u}^T B \bm{u}=1$ and $\bm{u}^Td>0$, we can choose the optimal value for the eigenvector $\bm{u}_{opt}$ as the scaled version of the eigenvector $\bm{u}_{max}$ corresponding to the eigenvalue $\lambda_{max}$:
\eqal{
    \bm{u}_{opt}(\gamma) = \frac{\text{sign}(\bm{u}_{max}^T\bm{d})}{\sqrt{\bm{u}_{max}^T B \bm{u}_{max}}} \bm{u}_{max}
}

\subsection{Subproblem 3: Derivation of $Q_3$} \label{appx:step3_deriveq3}
We know that the cost function $Q_3(\gamma)$ is the maximum eigenvalue of the GEP $\lambda_{max}\bigb{C_0 + \gamma C_1 + \gamma^2 C_2,B}$. An analytical expression of $\lambda_{max}$ as a function of $\gamma$ can be derived as follows:
\begin{itemize}
    \item All eigenvalues of the GEP must satisfy the characteristic equation:
    \eqal{        
        \det\bigb{C_0 + \gamma C_1 + \gamma^2 C_2 - \lambda B} = 0
    }
    \item Since $D_0$ contains a column of zeros, certain entries of $C_1$ and $C_2$ are also zero:
    \eqal{        
        \det\bigb{\mx{\times&\times\\\times&\times} + \gamma \mx{\times&\times\\\times&0} + \gamma^2 \mx{\times&0\\0&0} - \lambda \mx{0&2\\2&0}} = 0
    }
    where $\times$ denotes non-zero elements.
    \item Expressing the above determinant in scalar form, the following equation is obtained:
    \eq{ \label{eq:char_pol}
        4\lambda^2 + (t_1+\gamma t_2)\lambda + \gamma^2 t_3 + \gamma t_4 + t_5 = 0
    }
    where
    \eqal{
        t_1 &= 2(C_{0,12} + C_{0,21}) \\
        t_2 &= 2(C_{1,12} + C_{1,21}) \\
        t_3 &= C_{0,22}C_{2,11}-C_{1,12}C_{1,21} \\ 
        t_4 &= C_{0,22}C_{1,11}-C_{0,12}C_{1,21}-C_{0,21}C_{1,12} \\
        t_5 &= C_{0,11}C_{0,22}-C_{0,12}C_{0,21}
    }
    and $C_{,ij}$ denotes the element in the $i^{th}$ row and $j^{th}$ column of the respective matrix.
    \item Eq.~\eqref{eq:char_pol} contains a second-order polynomial in terms of $\lambda$, thus $\lambda_{max}$ can be analytically expressed as:
    \eqal{
        \lambda_{max}(\gamma)= \frac{-(t_1+\gamma t_2) + \sqrt{(t_1+\gamma t_2)^2-16(\gamma^2 t_3 + \gamma t_4 + t_5)}}{8}
    }
\end{itemize}

\subsection{Subproblem 3: Solution for $\gamma^*$} \label{appx:step3_gamma}
To identify the $\gamma$ value that minimizes $Q_3(\gamma)$, we set the derivative of $Q_3(\gamma)$ with respect to $\gamma$ to zero. Note that this results in a second-order polynomial equation in terms of $\gamma$ that has two roots:
\eq{ \label{eq:gamma12}
    \gamma_{1,2} = \frac{-(t_1t_2-8t_4) \pm \sqrt{(t_1t_2-8t_4)^2-4(t_2^2-16t_3)(t_1t_2t_4-t_2^2t_5-4t_4^2)/t_3} }{t_2^2-16t_3}
}
Since we aim to minimize $Q_3(\gamma)$, we chose the root that yields the lower cost: $\gamma^*= \argmin{\gamma \in \{\gamma_1,\gamma_2\}} Q_3(\gamma)$.

\subsection{Singularity detection} \label{appx:sing}
Voxel-wise singularity detection is performed based on the angle of inclination of the line that passes through both the rotated ellipse center and the midpoint of $\pi$-separated rotated signal pairs in the complex plane at N=4. This inclination angle can be analytically computed via the following expression: 
\eq{\Phi_{i} = \tan^{-1}\bigb{\frac{\frac{\tilde{y}_{i}+\tilde{y}_{i+2}}{2}}{\frac{\tilde{x}_{i}+\tilde{x}_{i+2}}{2}-x_c}}} where $i=\{1,2\}$. Here, $\tilde{x}_{i}$ and $\tilde{y}_{i}$ denote the real and imaginary components of back-rotated $i^{th}$ acquisition, $(\tilde{x},\tilde{y})_{i}$ and $(\tilde{x},\tilde{y})_{i+2}$ form a $\pi$-separated signal pair, and $x_c$ is the real component of the rotated ellipse center. In the noise-free case, two pairs of conjugate symmetric data points occur when $|\Phi_{1}| + |\Phi_{2}| = 0$. To account for noisy measurements, here voxels with $|\Phi_{1}| + |\Phi_{2}| < \pi/12$ were taken as singular cases. This specific singularity threshold was empirically observed to work robustly across broad ranges of tissue and sequence parameters. Note that singularities were detected in simulated experiments where it is more likely to exactly satisfy $|\Phi_{1}| + |\Phi_{2}| = 0$, yet no noticeable singularities were detected for phantom and in vivo experiments. 

\clearpage
\section{Supp. Text 2}\label{appx:seqs}

\textbf{Reference $\bm{T_1}$ mapping:} Reference $T_1$ maps for the phantom were collected using a 2D inversion recovery turbo spin echo sequence (IR-TSE) with inversion times $TI=\{25,50,100,200,500,1000,2000,4000\}$ ms, a turbo factor of $5$, a $TR/TE$ of $4500/11$ ms, an FOV of $175\text{ mm}\times175\text{ mm}$, a matrix size of $128\times128$. Total scan time was 16 min 24 s. Reference $T_1$ maps for the brain were collected using the same sequence, but with a turbo factor of $32$, a $TR/TE$ of $5000/7.1$ ms, an FOV of $256\text{ mm}\times256\text{ mm}$, a matrix size of $192\times192$. Total scan time was 4 min 56 s.
    
\textbf{Reference $\bm{T_2}$ mapping:} Reference $T_2$ maps for the phantom were collected using a 2D multi-echo spin echo (ME-SE) sequence with echo times $TE=\{12:12:240\}$ ms, a $TR$ of $5000$ ms, an FOV of $175\text{ mm}\times175\text{ mm}$, a matrix size of $128\times128$. Reference $T_2$ maps for the brain were collected using the same sequence, but with an FOV of $256\text{ mm}\times256\text{ mm}$, a matrix size of $128\times128$. Total scan times for both experiments were 10 min 47 s. All ME-SE $T_2$ maps were corrected with the EMC fit technique \cite{ben2015rapid}.

\textbf{$\bm{B_1}$ mapping:} $B_1$ field maps  were acquired using the Bloch-Siegert Shift method \cite{sacolick2010b1} with a $6$ $\mu s$ $1$ kHz Fermi Pulse, a $TR/TE$ of $100/15$, and with the same FOV and matrix size of corresponding bSSFP acquisitions. Total scan times per cross-section were 25 s for the phantom and 51 s for the in vivo experiment. Flip angle correction was applied voxelwise while estimating $T_1$ and $T_2$ values from Eq. (31) by replacing the nominal flip angle with the actual flip angle. 
    
\textbf{$\bm{B_0}$ mapping:} $B_0$ field maps were acquired with dual-echo gradient echo acquisitions with echo times $TE=\{5.19,7.65\}$ ms, a $TR$ of $400$ ms, a matrix size of $64\times64$, and identical FOV to corresponding bSSFP acquisitions. Total scan time was $50$ s. To check for potential changes in main field distribution due to $B_0$ drifts, field maps were acquired immediately prior to and after the set of N=8 bSSFP acquisitions. No major changes in the field were observed, thus correction for $B_0$ drift was omitted in the presented results. If a significant $B_0$ drift occurs, the correction procedure described in \cite{shcherbakova2019investigation} can be performed.

\clearpage
\section{Supp. Text 3}\label{appx:runtimes}

\textbf{Run times for the methods:} Parameter estimations were performed on a computer with Intel\textsuperscript{\tiny\textregistered} Core\textsuperscript{\tiny\texttrademark} i5-6500K CPU, 16 GB RAM. Run times of the proposed technique per cross-section are listed in Supp. Table \ref{tab:times}. There was negligible effect of N on run times, so averages run times are given across $N=4-8$. Dictionary construction is a one time process that is performed prior to processing data. Its run time depends on the size of the dictionary, which was 44.2 s in our case.

\newpage
\begin{table}[h]
\caption{$T_1$ and $T_2$ values of simulated tissues at 3T}
\centering 
\begin{tabular}{ccc} 
\hline\hline 
Tissue& $T_1$ (ms) & $T_2$ (ms) \\ [0.5ex]
\hline 
Fat & 350 & 130 \\ 
Bone marrow & 370 & 50 \\
Liver & 800 & 40 \\ 
White Matter & 1000 & 80 \\
Myocardium & 1150 & 45 \\
Vessels & 1200 & 50\\
Gray Matter & 1300 & 110  \\
Muscle & 1400 & 30 \\
CSF & 4000 & 1000 \\ 
\hline 
\end{tabular}
\label{tab:tissues}
\end{table}

\begin{table}[h]
\caption{Run times of CELF} \label{tab:run_times}
\centering 
\begin{tabular}{cccc} 
\hline\hline 
Experiment & Ellipse fit & Identification & Total \\ [0.5ex]
\hline 
Phantom & 1.5 s & 13.1 s & 14.6 s \\ 
In vivo & 3.6 s & 41.5 s & 45.1 s \\
\hline 
\end{tabular}
\label{tab:times}
\end{table}

\begin{table}[t]
\caption{Estimated $T_1$ and $T_2$ values for phantom \label{tab:phantom_roi}}
\centering
\begin{tabular}{|c|c|c|c|} 
\hline 
\multicolumn{2}{|c|}{\multirow{2}{*}{\textbf{Methods}}} & \multicolumn{2}{c|}{\textbf{Phantom}} \\ \cline{3-4} 
\multicolumn{2}{|c|}{} & $T_1$ (ms) & $T_2$ (ms) \\ \hline
\multirow{3}{*}{\rotatebox[origin=c]{0}{CELF}} & N=8 & 92.1 $\pm$ 5.7 & 80.6 $\pm$ 4.2 \\\cline{2-4}
& N=6 & 90.9 $\pm$ 4.1 & 77.5 $\pm$ 3.4 \\\cline{2-4}
& N=4 & 90.5 $\pm$ 5.1 & 76.1 $\pm$ 4.0 \\\hline
\multirow{2}{*}{\rotatebox[origin=c]{0}{PLANET}} & N=8 & 92.9 $\pm$ 6.5 & 81.3 $\pm$ 5.5 \\\cline{2-4}
& N=6 & 94.5 $\pm$ 8.4 & 79.7 $\pm$ 8.4 \\\hline
\multicolumn{2}{|c|}{Reference} & 100.5 $\pm$ 3.0 & 76.0 $\pm$ 0.9 \\\hline 
\end{tabular}
\end{table}

\begin{table}[t]
\caption{Estimated $T_1$ and $T_2$ values for in vivo brain images of S2 \label{tab:invivo_roi_subj2}}
\centering
\resizebox{\columnwidth}{!}{ 
\begin{tabular}{|c|c|c|c|c|c|c|c|} 
\hline 
\multicolumn{2}{|c|}{\multirow{2}{*}{\textbf{Methods}}} & \multicolumn{2}{c|}{\textbf{WM}} & \multicolumn{2}{c|}{\textbf{GM}} & \multicolumn{2}{c|}{\textbf{CSF}} \\ \cline{3-8} 
\multicolumn{2}{|c|}{} & $T_1$ (ms) & $T_2$ (ms) & $T_1$ (ms) & $T_2$ (ms) & $T_1$ (ms) & $T_2$ (ms) \\ \hline
\multirow{3}{*}{\rotatebox[origin=c]{0}{CELF}} & N=8 & 385 $\pm$ 69 & 55 $\pm$ 12 & 659 $\pm$ 161 & 75 $\pm$ 23 & 2350 $\pm$ 747 & 1174 $\pm$ 413 \\\cline{2-8}
& N=6 & 412 $\pm$ 81 & 51 $\pm$ 11 & 696 $\pm$ 238 & 73 $\pm$ 26 & 2512 $\pm$ 824 & 1210 $\pm$ 412 \\\cline{2-8}
& N=4 & 418 $\pm$ 94 & 56 $\pm$ 14 & 724 $\pm$ 187 & 81 $\pm$ 30 & 2707 $\pm$ 825 & 1231 $\pm$ 391 \\\hline
\multirow{2}{*}{\rotatebox[origin=c]{0}{PLANET}} & N=8 & 438 $\pm$ 152 & 58 $\pm$ 17 & 859 $\pm$ 383 & 89 $\pm$ 42 & - & - \\\cline{2-8}
& N=6 & 429 $\pm$ 159 & 59 $\pm$ 19 & 827 $\pm$ 348 & 89 $\pm$ 47 & - & - \\\hline
\multicolumn{2}{|c|}{Reference} & 545 $\pm$ 17 & 56 $\pm$ 3 & 1168 $\pm$ 50 & 81 $\pm$ 14 & 3734 $\pm$ 1080 & 825 $\pm$ 227 \\\hline 
\end{tabular}
}
\end{table}

\begin{table}[t]
\caption{Estimated $T_1$ and $T_2$ values for in vivo brain images of S3 \label{tab:invivo_roi_subj3}}
\centering
\resizebox{\columnwidth}{!}{ 
\begin{tabular}{|c|c|c|c|c|c|c|c|} 
\hline 
\multicolumn{2}{|c|}{\multirow{2}{*}{\textbf{Methods}}} & \multicolumn{2}{c|}{\textbf{WM}} & \multicolumn{2}{c|}{\textbf{GM}} & \multicolumn{2}{c|}{\textbf{CSF}} \\ \cline{3-8} 
\multicolumn{2}{|c|}{} & $T_1$ (ms) & $T_2$ (ms) & $T_1$ (ms) & $T_2$ (ms) & $T_1$ (ms) & $T_2$ (ms) \\ \hline
\multirow{3}{*}{\rotatebox[origin=c]{0}{CELF}} & N=8 & 334 $\pm$ 49 & 52 $\pm$ 9 & 722 $\pm$ 130 & 66 $\pm$ 16 & 2891 $\pm$ 1030 & 1080 $\pm$ 765 \\\cline{2-8}
& N=6 & 377 $\pm$ 82 & 49 $\pm$ 9 & 839 $\pm$ 377 & 64 $\pm$ 19 & 3076 $\pm$ 1033 & 1146 $\pm$ 788 \\\cline{2-8}
& N=4 & 344 $\pm$ 66 & 53 $\pm$ 10 & 767 $\pm$ 196 & 65 $\pm$ 20 & 2457 $\pm$ 1352 & 830 $\pm$ 649 \\\hline
\multirow{2}{*}{\rotatebox[origin=c]{0}{PLANET}} & N=8 & 402 $\pm$ 104 & 54 $\pm$ 9 & 736 $\pm$ 242 & 65 $\pm$ 16 & - & - \\\cline{2-8}
& N=6 & 374 $\pm$ 97 & 53 $\pm$ 12 & 741 $\pm$ 406 & 64 $\pm$ 19 & - & - \\\hline
\multicolumn{2}{|c|}{Reference} & 516 $\pm$ 17 & 63 $\pm$ 5 & 867 $\pm$ 187 & 61 $\pm$ 9 & 3086 $\pm$ 812 & 702 $\pm$ 275 \\\hline 
\end{tabular}
}
\end{table}

\begin{table}[t]
\caption{Estimated $T_1$ and $T_2$ values for in vivo brain images of S4 \label{tab:invivo_roi_subj4}}
\centering
\resizebox{\columnwidth}{!}{ 
\begin{tabular}{|c|c|c|c|c|c|c|c|} 
\hline 
\multicolumn{2}{|c|}{\multirow{2}{*}{\textbf{Methods}}} & \multicolumn{2}{c|}{\textbf{WM}} & \multicolumn{2}{c|}{\textbf{GM}} & \multicolumn{2}{c|}{\textbf{CSF}} \\ \cline{3-8} 
\multicolumn{2}{|c|}{} & $T_1$ (ms) & $T_2$ (ms) & $T_1$ (ms) & $T_2$ (ms) & $T_1$ (ms) & $T_2$ (ms) \\ \hline
\multirow{3}{*}{\rotatebox[origin=c]{0}{CELF}} & N=8 & 347 $\pm$ 62 & 57 $\pm$ 11 & 800 $\pm$ 186 & 64 $\pm$ 13 & 2933 $\pm$ 692 & 1369 $\pm$ 406 \\\cline{2-8}
& N=6 & 394 $\pm$ 71 & 52 $\pm$ 9 & 816 $\pm$ 164 & 62 $\pm$ 11 & 3048 $\pm$ 573 & 1215 $\pm$ 386 \\\cline{2-8}
& N=4 & 372 $\pm$ 68 & 54 $\pm$ 11 & 808 $\pm$ 223 & 62 $\pm$ 14 & 3096 $\pm$ 545 & 1284 $\pm$ 421 \\\hline
\multirow{2}{*}{\rotatebox[origin=c]{0}{PLANET}} & N=8 & 384 $\pm$ 77 & 58 $\pm$ 10 & 782 $\pm$ 3673 & 62 $\pm$ 19 & - & - \\\cline{2-8}
& N=6 & 385 $\pm$ 75 & 56 $\pm$ 11 & 1331 $\pm$ 2510 & 93 $\pm$ 23 & - & - \\\hline
\multicolumn{2}{|c|}{Reference} & 519 $\pm$ 24 & 64 $\pm$ 9 & 776 $\pm$ 154 & 55 $\pm$ 8 & 3333 $\pm$ 692 & 796 $\pm$ 231 \\\hline 
\end{tabular}
}
\end{table}

\begin{table}[t]
\caption{Discrepancy between CELF and reference $T_1$ and $T_2$ estimates before and after neural-network correction. Percentage difference is reported for each subject, along with mean and standard deviation across subjects ath the bottom row. \label{tab:invivo_NN_mape}}
\centering
\resizebox{\columnwidth}{!}{ 
\begin{tabular}{|c|c|c|c|c|c|c|c|} 
\hline 
\multicolumn{2}{|c|}{\multirow{2}{*}{\textbf{}}} & \multicolumn{2}{c|}{\textbf{WM}} & \multicolumn{2}{c|}{\textbf{GM}} & \multicolumn{2}{c|}{\textbf{CSF}} \\ \cline{3-8} 
\multicolumn{2}{|c|}{} & $T_1$ (\%) & $T_2$ (\%) & $T_1$ (\%) & $T_2$ (\%) & $T_1$ (\%) & $T_2$ (\%) \\ \hline
\multirow{2}{*}{\rotatebox[origin=c]{0}{S1}} & Before & 33.9 & 19.9 & 2.3 & 10.7 & 21.3 & 111.9 \\\cline{2-8}
& After & 2.4 & 7.6 & 1.7 & 11.3 & 26.4 & 33.3 \\\hline
\multirow{2}{*}{\rotatebox[origin=c]{0}{S2}} & Before & 29.5 & 1.3 & 43.6 & 6.3 & 26.0 & 70.3 \\\cline{2-8}
& After & 11.4 & 6.8 & 28.3 & 21.9 & 32.3 & 44.5 \\\hline
\multirow{2}{*}{\rotatebox[origin=c]{0}{S3}} & Before & 35.4 & 18.1 & 16.7 & 7.5 & 6.3 & 53.9 \\\cline{2-8}
& After & 9.2 & 5.1 & 0.1 & 2.5 & 30.6 & 46.7 \\\hline
\multirow{2}{*}{\rotatebox[origin=c]{0}{S4}} & Before & 33.1 & 10.6 & 3.1 & 17.1 & 12.0 & 72.1 \\\cline{2-8}
& After & 13.2 & 7.3 & 17.3 & 11.8 & 22.3 & 39.5 \\\hline
\multirow{2}{*}{\rotatebox[origin=c]{0}{All}} & Before & 33.0 $\pm$ 2.5 & 12.5 $\pm$ 8.5 & 16.4 $\pm$ 19.3 & 10.4 $\pm$ 4.8 & 16.4 $\pm$ 8.9 & 77.0 $\pm$ 24.6 \\\cline{2-8}
& After & 9.0 $\pm$ 4.7 & 6.7 $\pm$ 1.1 & 11.8 $\pm$ 13.4 & 11.9 $\pm$ 7.9 & 27.9 $\pm$ 4.5 & 41.0 $\pm$ 6.0 \\\hline
\end{tabular}
}
\end{table}

\begin{figure*}[t]
\centering
\includegraphics[width=\textwidth]{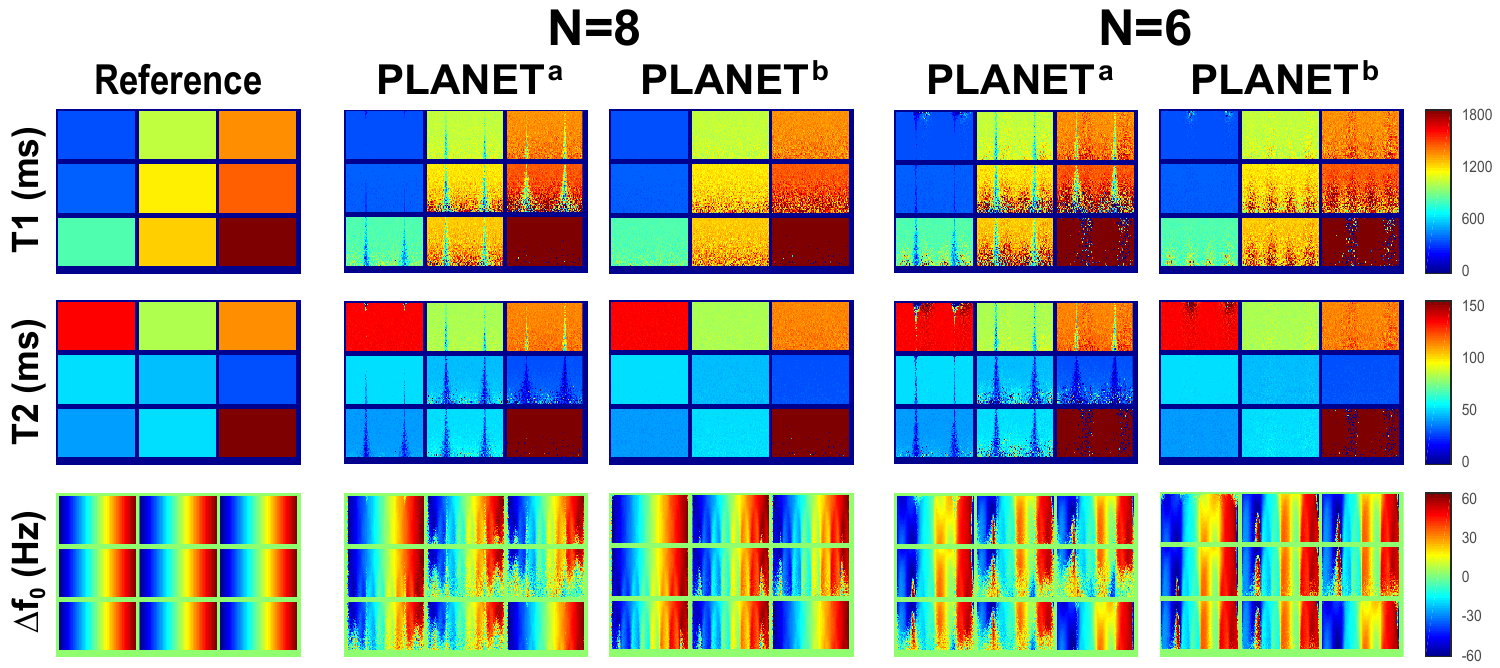}
\caption{\label{fig:sim2_planet} $T_1$, $T_2$, and off-resonance estimations of two variants of PLANET: back-rotation with $\varphi_{rot}$ (PLANET$^a$), and with $\phi$ (PLANET$^b$) for various tissues (left column from top to bottom: fat, bone marrow, liver; middle column from top to bottom: white matter, myocardium, vessels; right column from top to bottom: gray matter, muscle, and CSF). Each small rectangle has varying off-resonance and flip angle values; from left to right off-resonance changes -62.5 Hz to 62.5 Hz ($0.5/TR$), from top to bottom flip angle changes 20$^o$ to 60$^o$.}
\end{figure*}

\begin{figure}[t]
\centering
\includegraphics[width=\columnwidth]{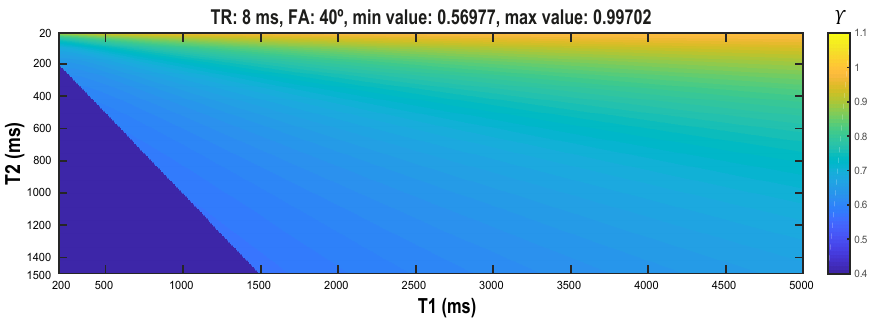}
\caption{\label{fig:gamma} To determine the possible range of $\gamma$, bSSFP signal parameters were simulated. Representative results are displayed for a TR of 8 ms, a flip angle of $40^o$, $T_1 \in [200 \mbox{ } 5000]$ ms and $T_2 \in [10 \mbox{ } 1500]$ ms such that $T_1 \ge T_2$.}
\end{figure}

\begin{figure*}[t]
\centering
\includegraphics[width=\columnwidth]{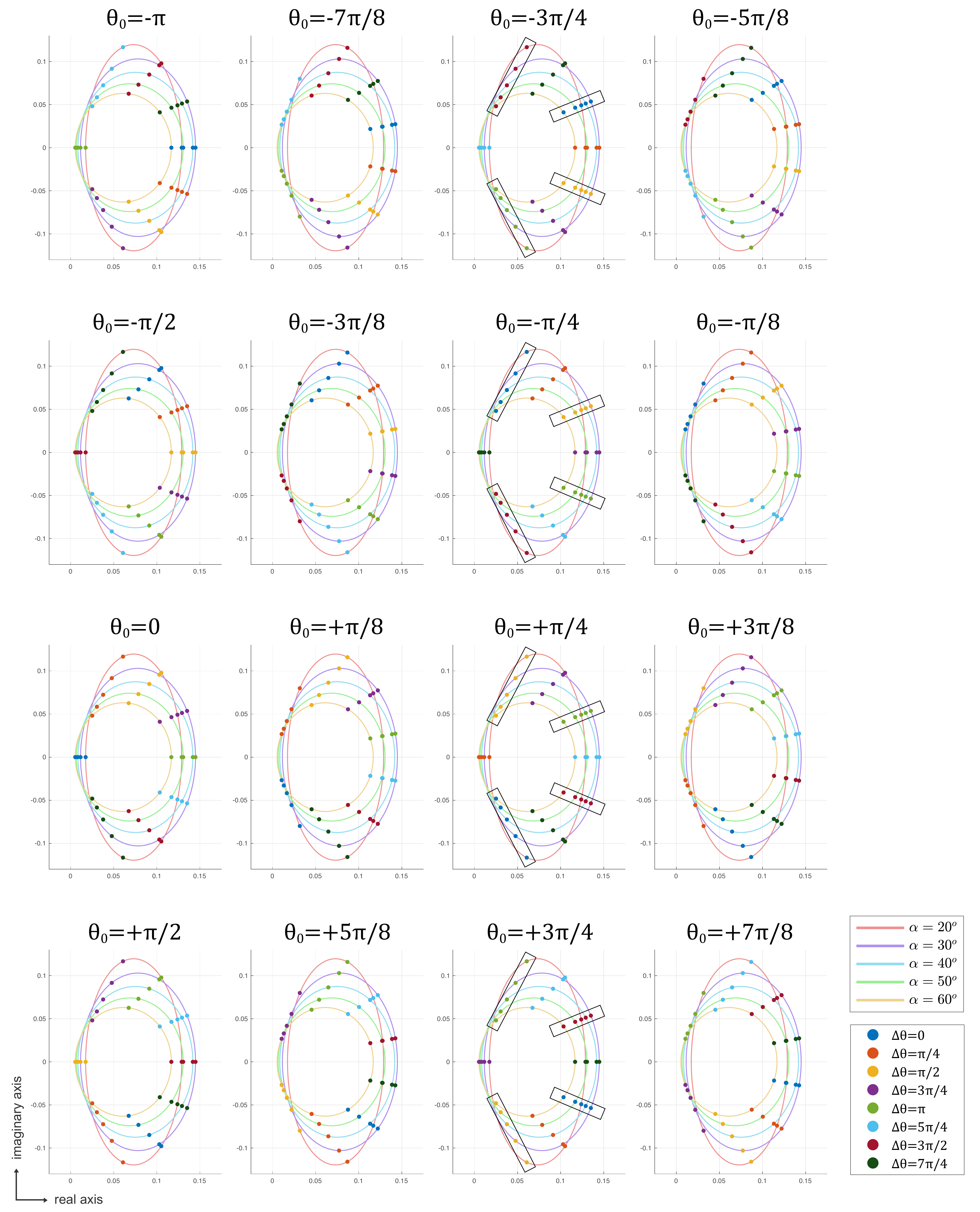}
\caption{\label{fig:sim2_sing_ellipse_supp} Phase-cycled bSSFP signals were simulated for gray matter, and the resulting vertical bSSFP ellipses were displayed in the complex plane for a separate $\theta_0$ value in each plot. In each plot, ellipses for different flip angles ($\alpha=\{20^o, 30^o, 50^o, 60^o\}$) are marked with different colors (see top legend). For each ellipse, $N=8$ measurements obtained with phase-cycling increments at $\Delta\theta=\{0,\pi/4,\pi/2,3\pi/4,\pi,5\pi/4,3\pi/2,7\pi/4\}$ are marked with colored circles (see bottom legend). Note that singularities occur at specific $\theta_0$ values when $N=4$. For each plot, the set of 4 measurements (at $\theta=\{0,\pi/2,\pi,3\pi/2\}$) that suffer from a singularity are enclosed within black rectangles. A singularity occurs when measurements are symmetric around the real axis. This also means that if another set of $\Delta\theta$ values were chosen with $N=4$, $\theta_0$ values at which singularities are observed would be altered. For example, if $N=4$ acquisitions are performed at $\Delta\theta=\{\pi/4,3\pi/4,5\pi/4,7\pi/4\}$, symmetric measurements, therefore singularities, will occur for $\theta_0=\{-\pi,-\pi/2,0,+\pi/2\}$.}
\end{figure*}

\begin{figure*}[t]
\centering
\includegraphics[width=0.6\textwidth]{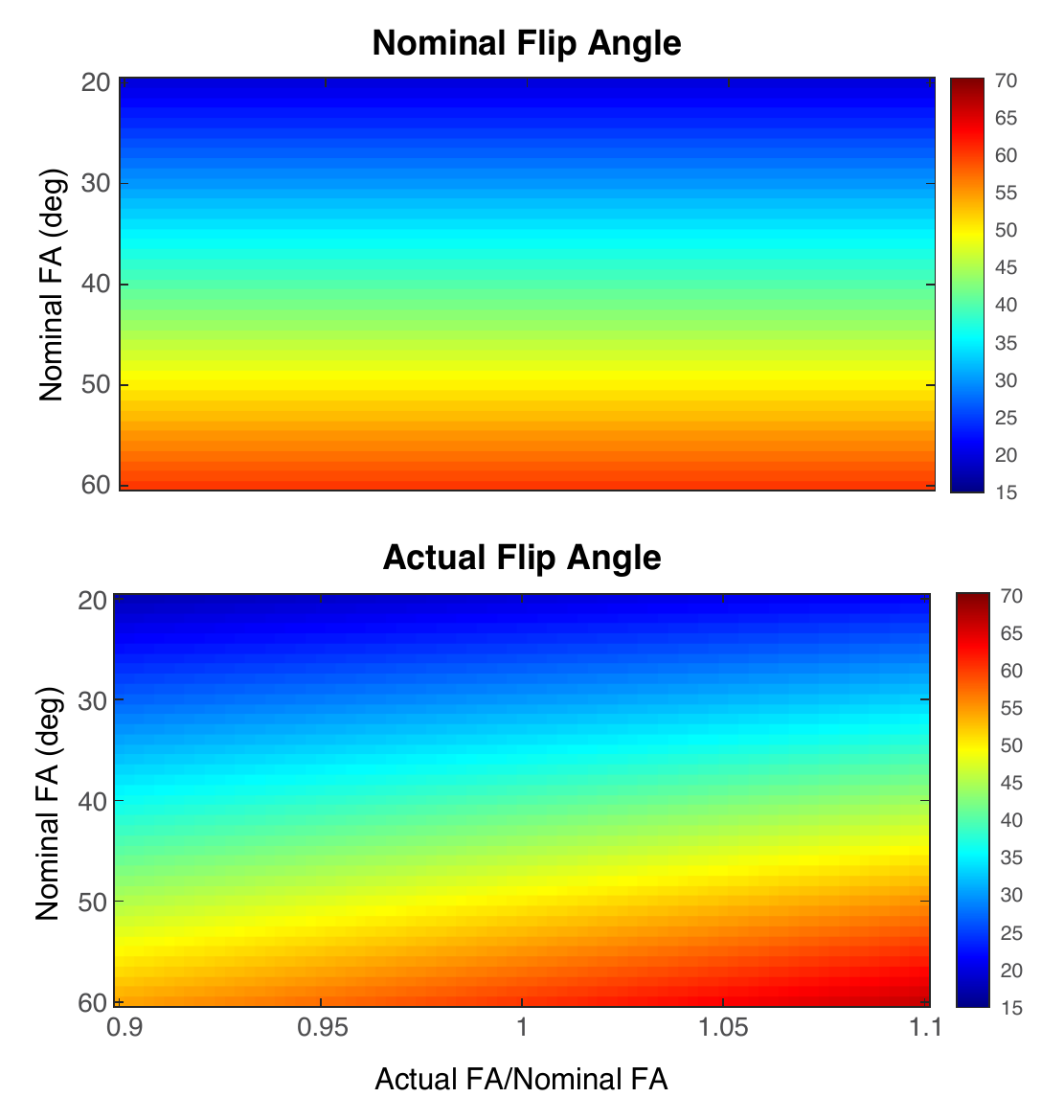}
\caption{\label{fig:sim2_FA_deg_supp} Nominal and actual flip-angle maps generated across tissue blocks.}
\end{figure*}

\begin{figure*}[t]
\centering
\includegraphics[width=0.9\textwidth]{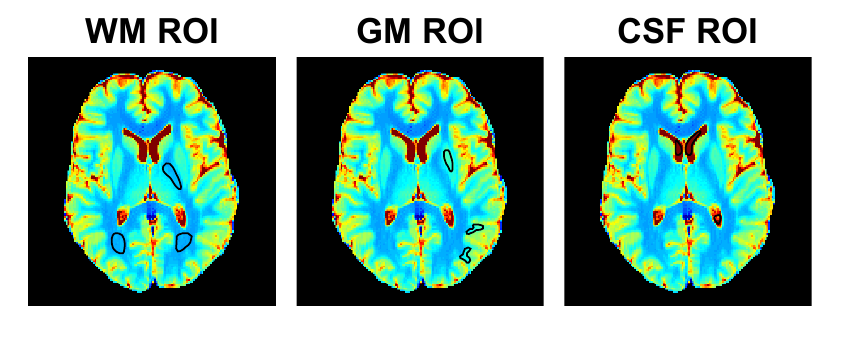}
\caption{\label{fig:invivo_subj1_ROI_supp} Parameter estimates for specific tissues were examined on manually-defined ROIs. White matter (WM), gray matter (GM), and CSF ROIs are shown for a representative cross-section in S1. Total number of voxels inside the ROIs were $346$, $119$ and $171$, respectively. ROIs are encircled with solid black lines on the reference $T_1$ map.}
\end{figure*}

\begin{figure*}[t]
\centering
\includegraphics[width=0.9\textwidth]{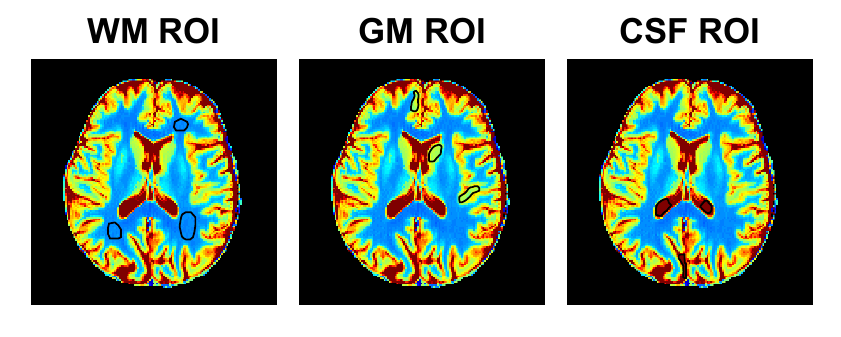}
\caption{\label{fig:invivo_subj2_ROI_supp} Parameter estimates for specific tissues were examined on manually-defined ROIs. White matter (WM), gray matter (GM), and CSF ROIs are shown for a representative cross-section in S2. Total number of voxels inside the ROIs were $577$, $269$ and $193$, respectively. ROIs are encircled with solid black lines on the reference $T_1$ map.}
\end{figure*}

\begin{figure*}[t]
\centering
\includegraphics[width=\textwidth]{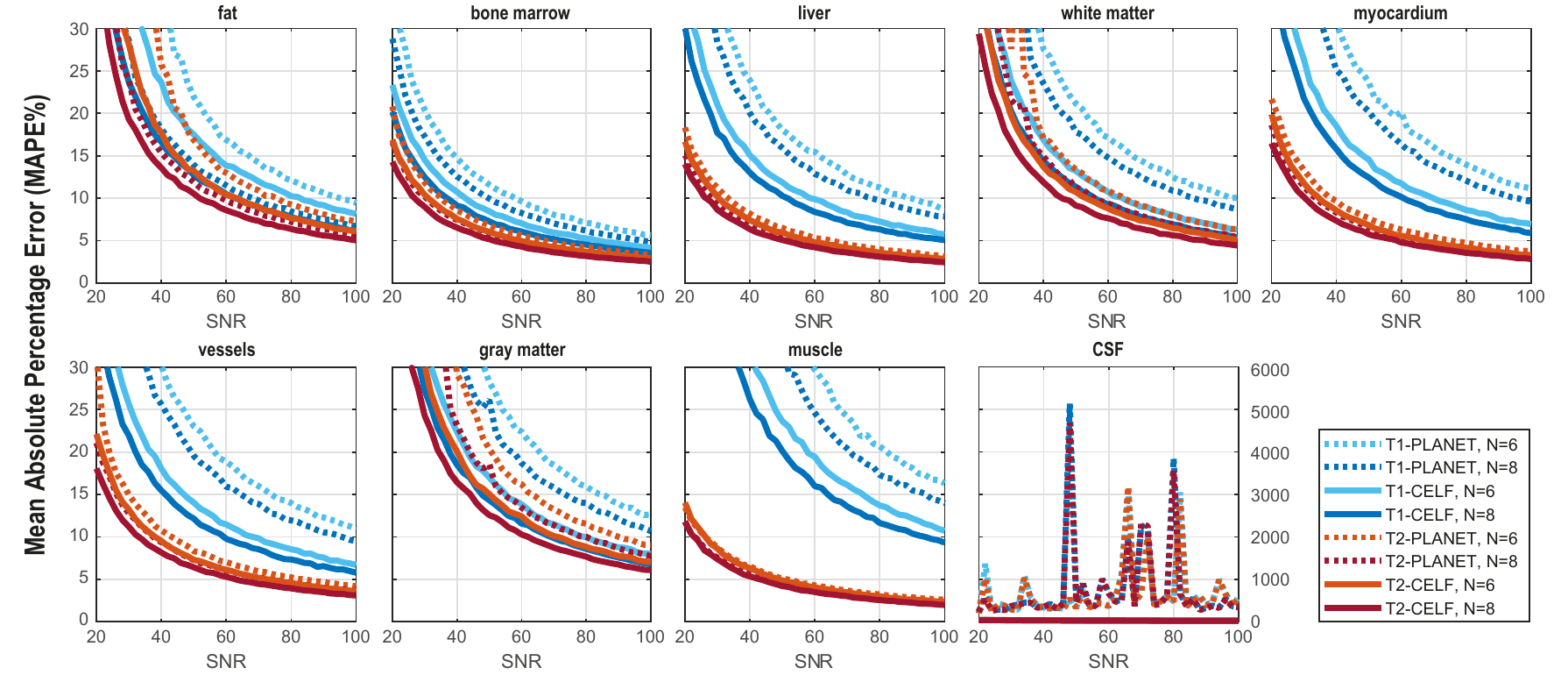}
\caption{\label{fig:sim_mape} Phase-cycled bSSFP signals were simulated for nine tissues (fat, bone marrow, liver, white matter, myocardium, vessels, gray matter, muscle, and CSF) and for SNR levels varying in [20 100] in each tissue. The remaining simulation parameters were: $TR=8$ms, $TE=4$ms, flip angle $\alpha=40^o$ and $N=\{6,8\}$. CELF and PLANET were employed to estimate relaxation parameters. Errors for $T_1$ estimation are marked with blue lines, those for $T_2$ estimation are marked with red lines. Meanwhile, CELF results are shown with solid lines and PLANET results are shown with dashed lines.}
\end{figure*}

\begin{figure*}[t]
\centering
\includegraphics[width=\textwidth]{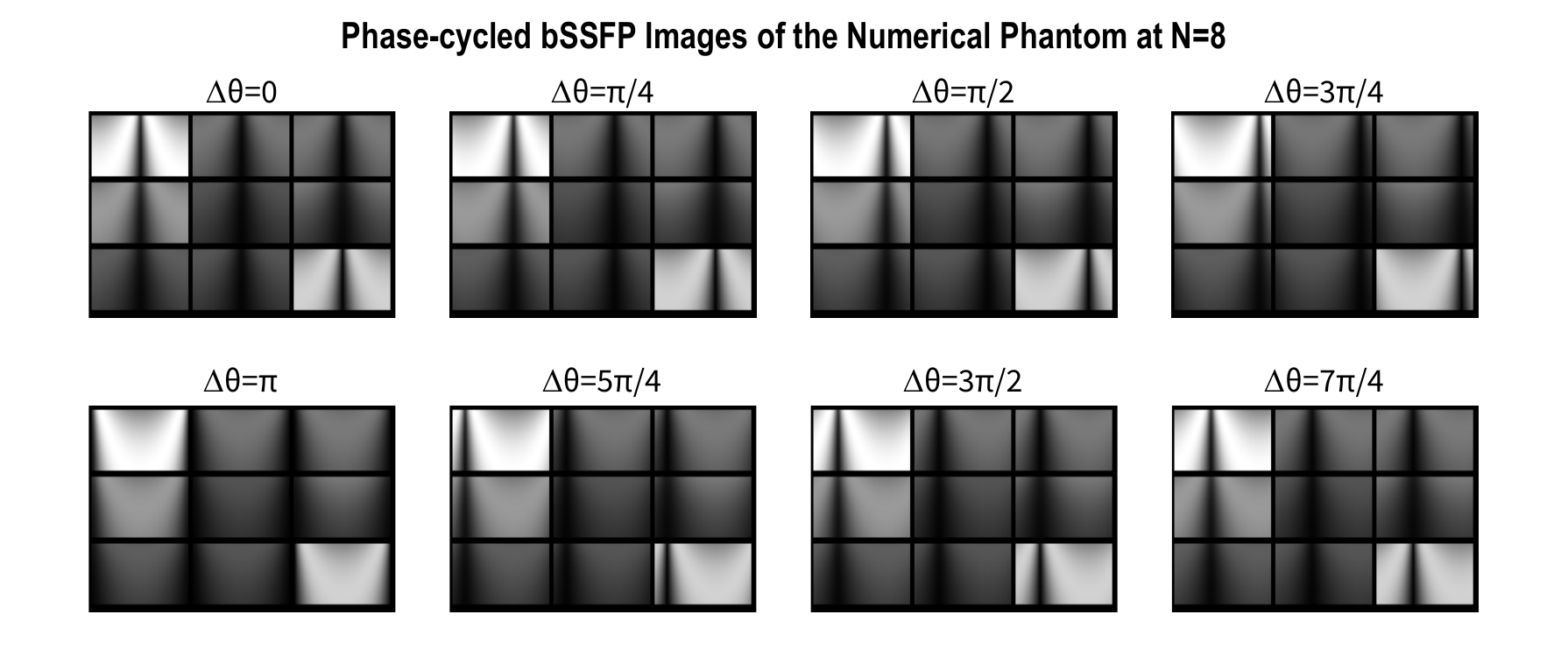}
\caption{\label{fig:sim2_pc_supp} Phase-cycled bSSFP signals were simulated for nine tissue blocks (left column from top to bottom: fat, bone marrow, liver; middle column from top to bottom: white matter, myocardium, vessels; right column from top to bottom: gray matter, muscle, and CSF). In each block off-resonance varied from -62.5 Hz to 62.5 Hz ($0.5/TR$) along the horizontal axis, and flip angle varied from 20$^o$ to 60$^o$ along the vertical axis. Phase-cycled bSSFP images of this numerical phantom at $N=8$ with $\Delta\theta=\{0,\pi/4,3\pi/4,\pi/2,5\pi/4,3\pi/2,7\pi/4\}$ are shown. Remaining parameters were: $TR=8$ms, $TE=4$ms, $N=\{4,6,8\}$, and SNR = 200 (with respect to CSF signal intensity).}
\end{figure*}

\begin{figure*}[t]
\centering
\includegraphics[width=\textwidth]{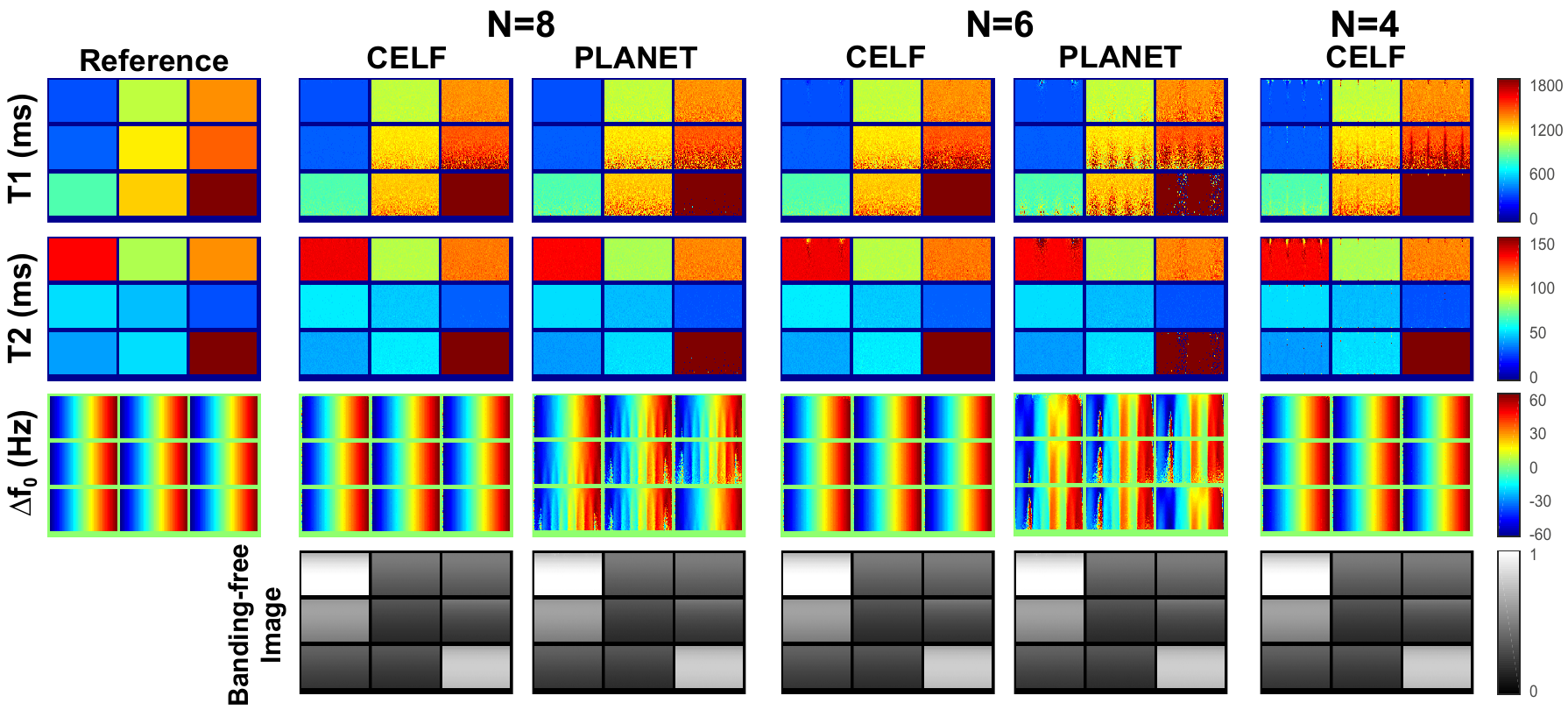}
\caption{\label{fig:sim2_supp} Phase-cycled bSSFP signals were simulated for nine tissue blocks (left column from top to bottom: fat, bone marrow, liver; middle column from top to bottom: white matter, myocardium, vessels; right column from top to bottom: gray matter, muscle, and CSF). In each block off-resonance varied from -62.5 Hz to 62.5 Hz ($0.5/TR$) along the horizontal axis, and flip angle varied from 20$^o$ to 60$^o$ along the vertical axis. Remaining parameters were: $TR=8$ms, $TE=4$ms, $N=\{4,6,8\}$, and SNR = 200 (with respect to CSF signal intensity). $T_1$, $T_2$, off-resonance, and banding-free image estimates via CELF and PLANET are displayed.}
\end{figure*}

\begin{figure*}[t]
\centering
\includegraphics[width=0.7\textwidth]{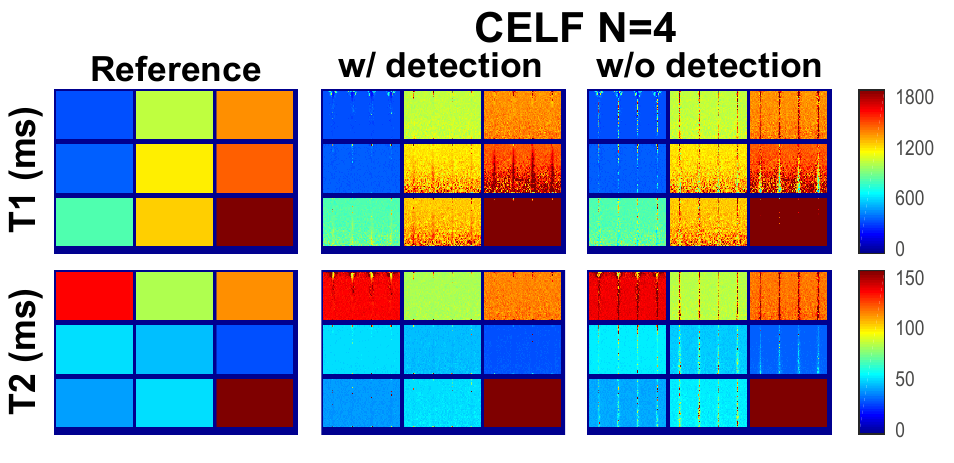}
\caption{\label{fig:sim2_supp_sing} Phase-cycled bSSFP signals were simulated for nine tissue blocks (left column from top to bottom: fat, bone marrow, liver; middle column from top to bottom: white matter, myocardium, vessels; right column from top to bottom: gray matter, muscle, and CSF). In each block off-resonance varied from -62.5 Hz to 62.5 Hz ($0.5/TR$) along the horizontal axis, and flip angle varied from 20$^o$ to 60$^o$ along the vertical axis. Remaining parameters were: $TR=8$ms, $TE=4$ms, $N=\{4,6,8\}$, and SNR = 200 (with respect to CSF signal intensity). $T_1$ and $T_2$ estimates via CELF with and without singularity detection are displayed. Off-resonance and banding-free estimates are not displayed here as they are not affected from singularities.}
\end{figure*}

\begin{figure*}[t]
\centering
\includegraphics[width=0.7\textwidth]{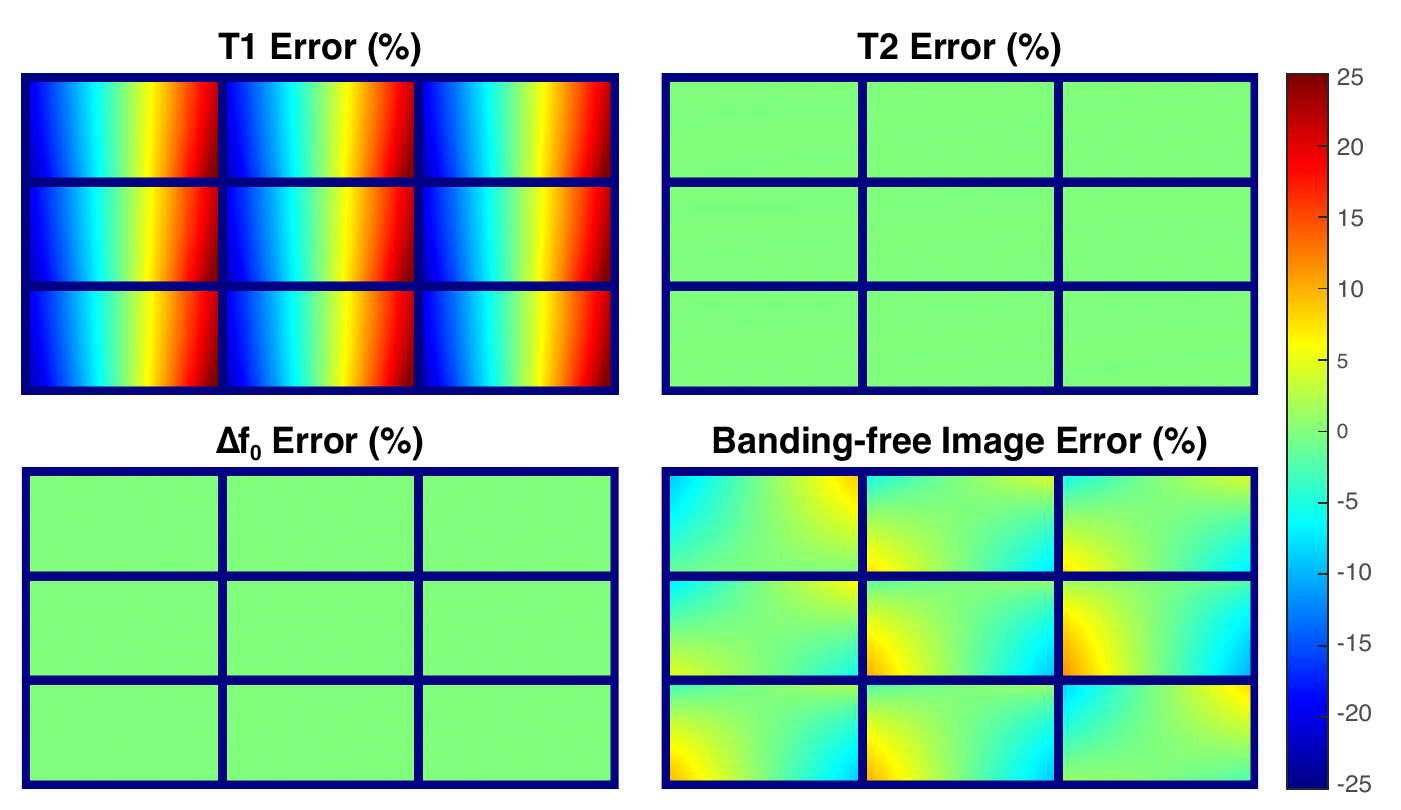}
\caption{\label{fig:sim2_FA_error_supp} Sensitivity of CELF to changes in flip angle variation are shown with phase-cycled bSSFP signals simulated in nine tissue blocks (left column from top to bottom: fat, bone marrow, liver; middle column from top to bottom: white matter, myocardium, vessels; right column from top to bottom: gray matter, muscle, and CSF). In each block nominal flip angle varied from 20$^o$ to 60$^o$ along the vertical axis, and the ratio of actual flip angle and nominal flip angle varied from 0.9 to 1.1 along the horizontal axis (see Supp. Fig. \ref{fig:sim2_FA_deg_supp} for the flip angle maps). Remaining parameters were: $TR=8$ms, $TE=4$ms, $N=4$. Percentage errors in $T_1$, $T_2$, off-resonance and banding-free image are displayed.}
\end{figure*}

\begin{figure*}[t]
\centering
\includegraphics[width=0.8\textwidth]{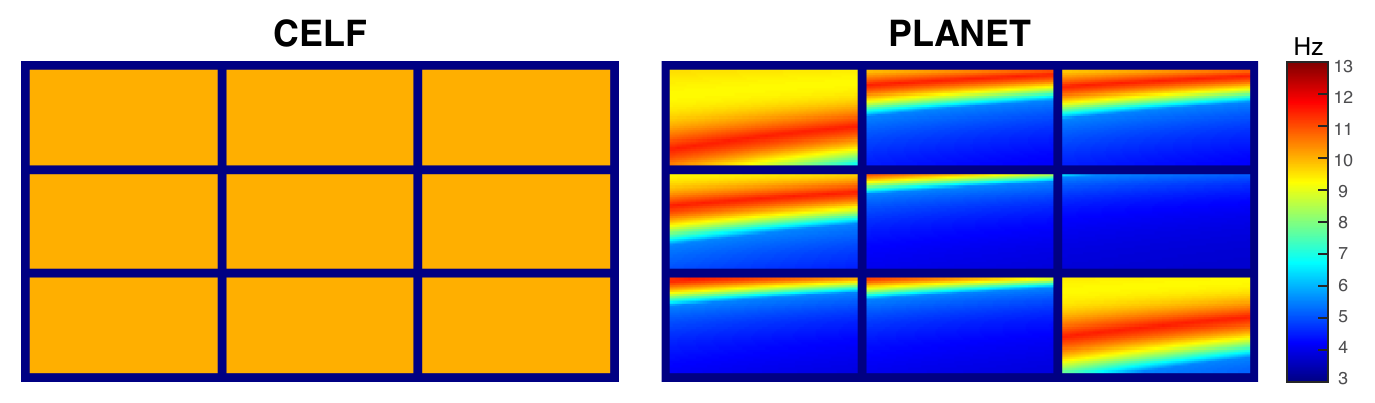}
\caption{\label{fig:sim2_FA_comp_supp} Sensitivity of off-resonance estimates from CELF and PLANET to flip angle variation are shown. Phase-cycled bSSFP signals were simulated in nine tissue blocks (left column from top to bottom: fat, bone marrow, liver; middle column from top to bottom: white matter, myocardium, vessels; right column from top to bottom: gray matter, muscle, and CSF). In each block nominal flip angle varied from 20$^o$ to 60$^o$ along the vertical axis, and the ratio of actual flip angle and nominal flip angle varied from 0.9 to 1.1 along the horizontal axis. Remaining parameters were: $TR=8$ms, $TE=4$ms, $N=4$, and off-resonance = 10 Hz. Off-resonance values estimated across each tissue block are shown.}
\end{figure*}

\begin{figure*}[t]
\centering
\includegraphics[width=\columnwidth]{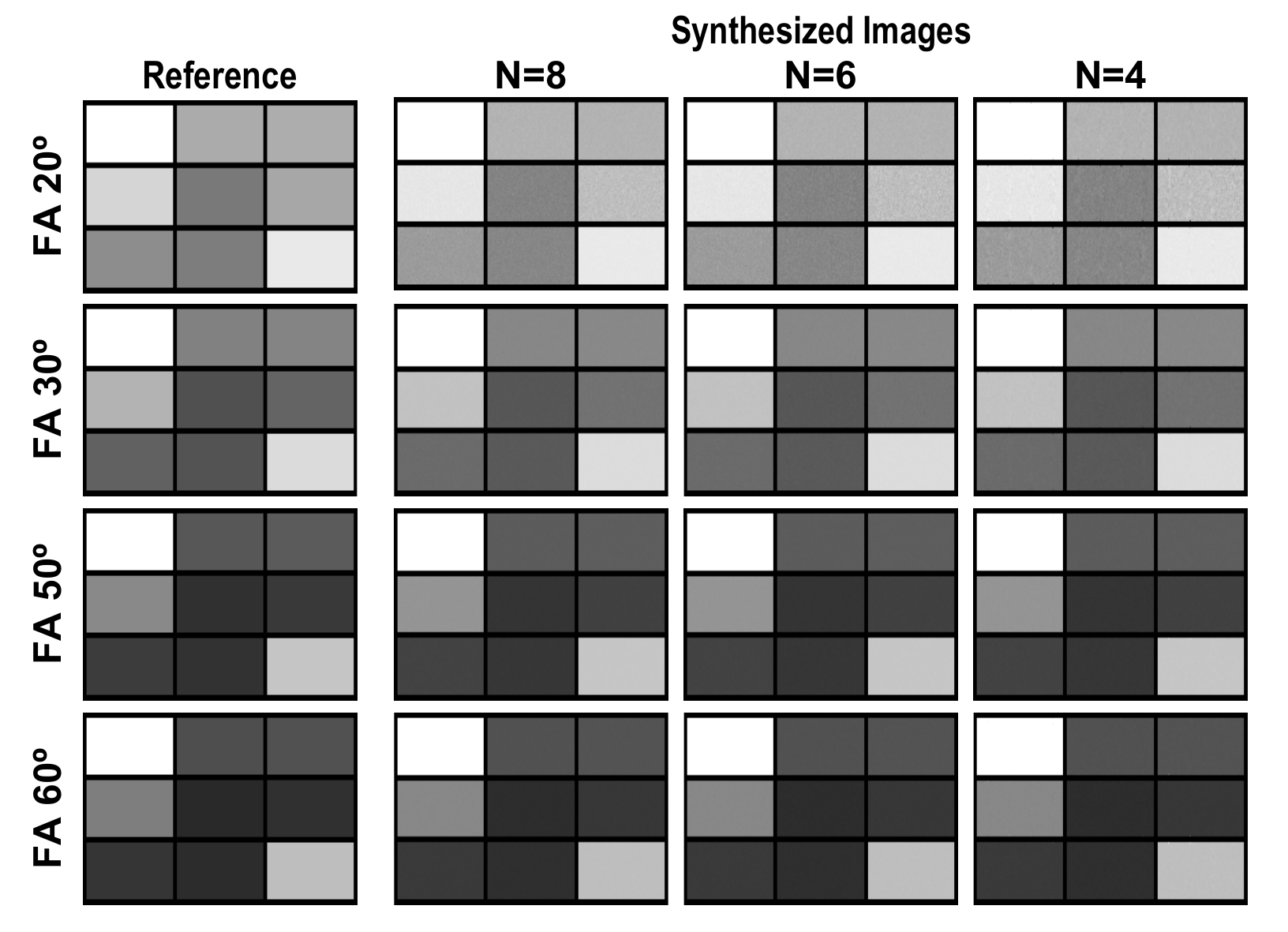}
\caption{\label{fig:sim2_FAsynt_supp} Synthetic bSSFP images at varying flip angles were generated based on CELF parameter estimates obtained a a specific flip angle. For demonstration, phase-cycled bSSFP signals were simulated for nine tissue blocks (left column from top to bottom: fat, bone marrow, liver; middle column from top to bottom: white matter, myocardium, vessels; right column from top to bottom: gray matter, muscle, and CSF). In each block off-resonance varied from -62.5 Hz to 62.5 Hz ($0.5/TR$) along the horizontal axis. Remaining parameters were: flip angle$=40^o$, $TR=8$ms, $TE=4$ms, $N=\{4,6,8\}$, and SNR = 200 (with respect to CSF signal intensity). First, $T_1$, $T_2$, and banding free image estimates at flip angle $=40^o$ were obtained with CELF. Next, CELF parameter estimates were used to synthesize banding-free bSSFP images at flip angles $\{20^o, 30^o, 50^o, 60^o\}$. The CELF-derived synthetic bSSFP images at $N=\{4,6,8\}$ are virtually identical to reference bSSFP images directly simulated for each flip angle.} 
\end{figure*}

\begin{figure*}[t]
\centering
\includegraphics[width=\textwidth]{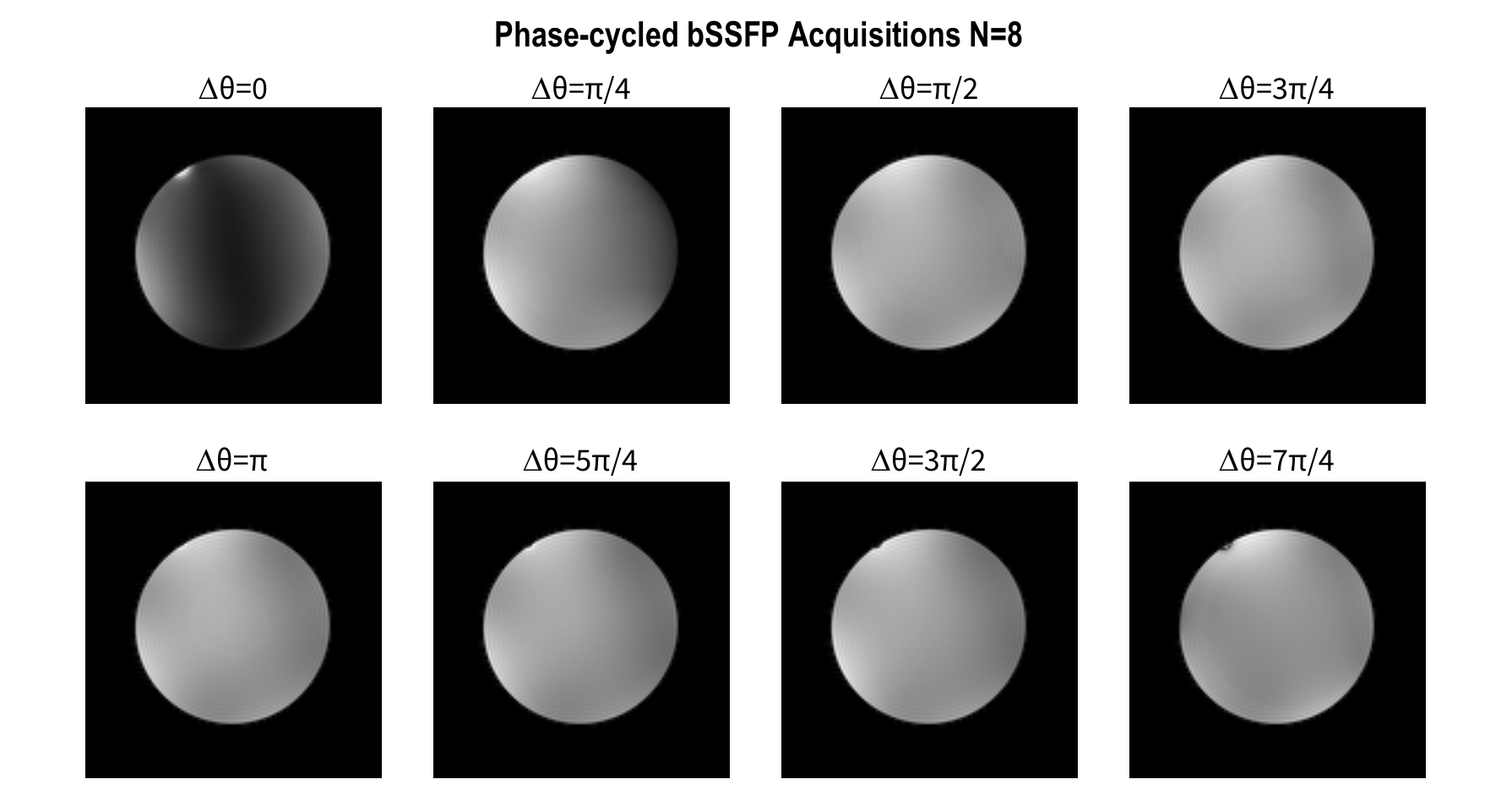}
\caption{\label{fig:phantom_pc_sup} Images of a phantom from phase-cycled bSSFP acquisitions with $\Delta\theta=\{0,\pi/4,3\pi/4,$ $\pi/2,5\pi/4,3\pi/2,7\pi/4\}$ are shown.}
\end{figure*}

\begin{figure*}[t]
\centering
\includegraphics[width=\textwidth]{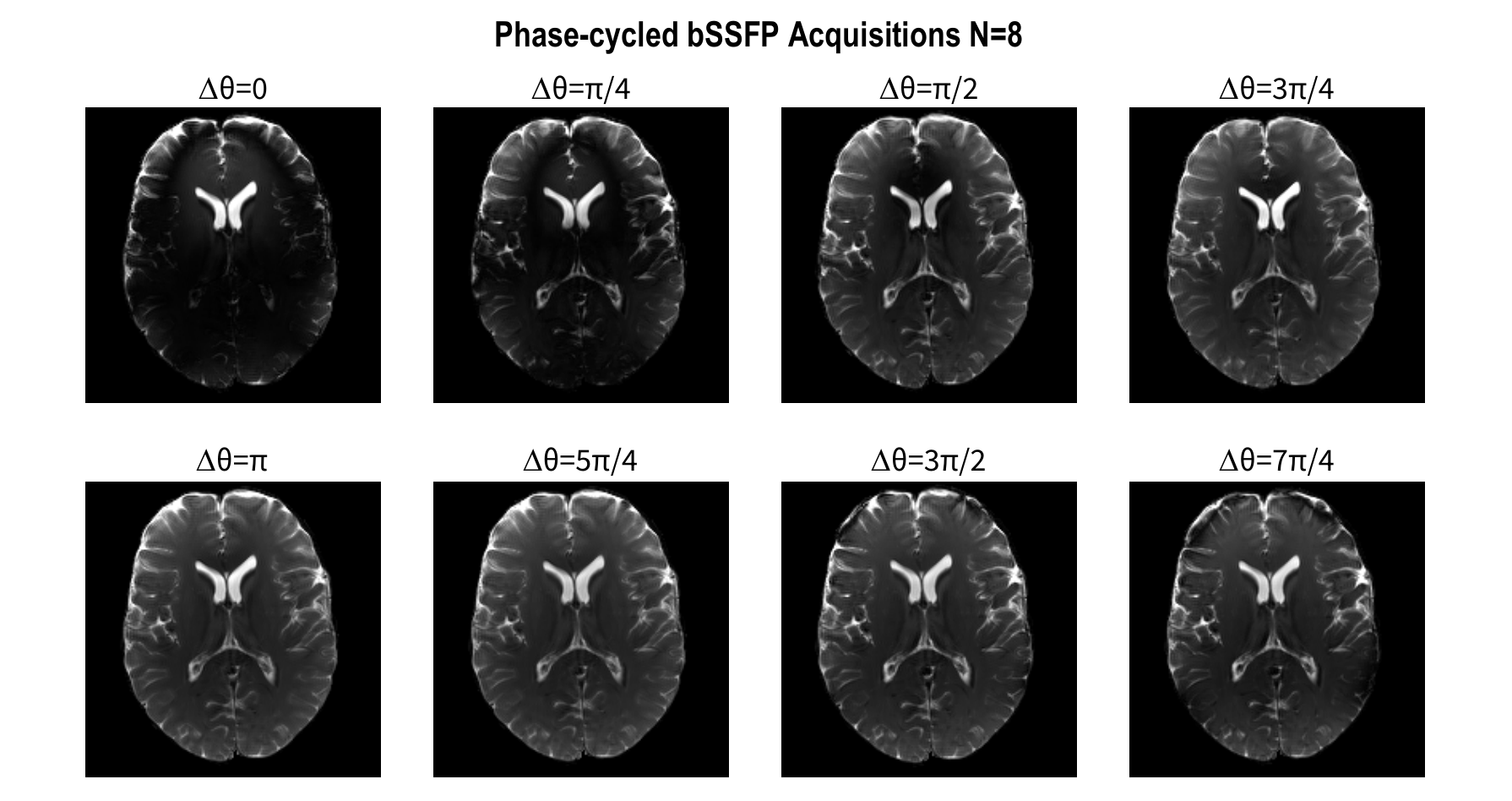}
\caption{\label{fig:invivo_pc_subj1_supp} In vivo brain images of S1 from phase-cycled bSSFP acquisitions with $\Delta\theta=\{0,\pi/4,3\pi/4,\pi/2,5\pi/4,3\pi/2,7\pi/4\}$ are shown.}
\end{figure*}

\begin{figure*}[t]
\centering
\includegraphics[width=\textwidth]{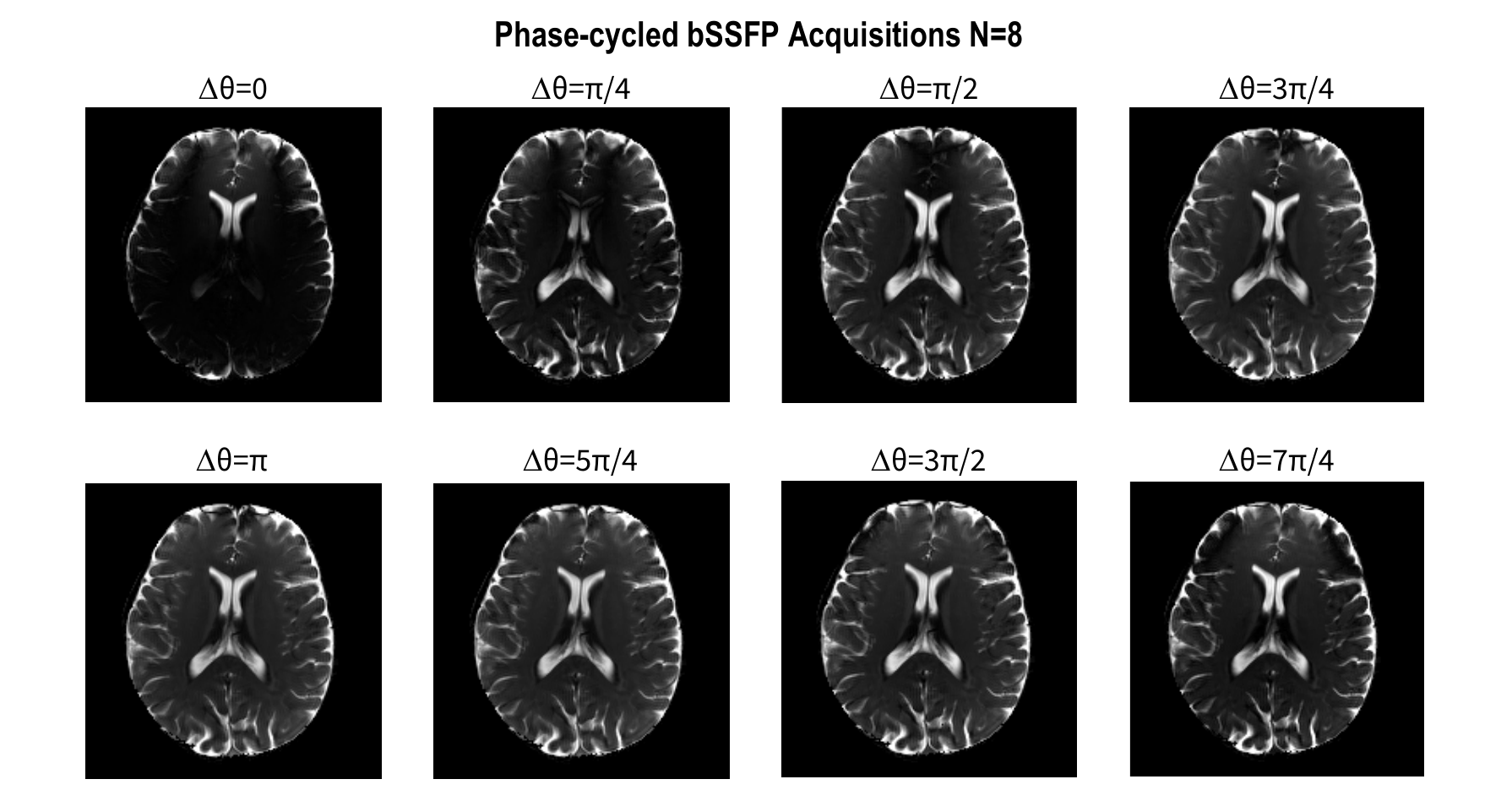}
\caption{\label{fig:invivo_pc_subj2_supp} In vivo brain images of S2 from phase-cycled bSSFP acquisitions with $\Delta\theta=\{0,\pi/4,3\pi/4,\pi/2,5\pi/4,3\pi/2,7\pi/4\}$ are shown.}
\end{figure*}

\begin{figure*}[t]
\centering
\includegraphics[width=\textwidth]{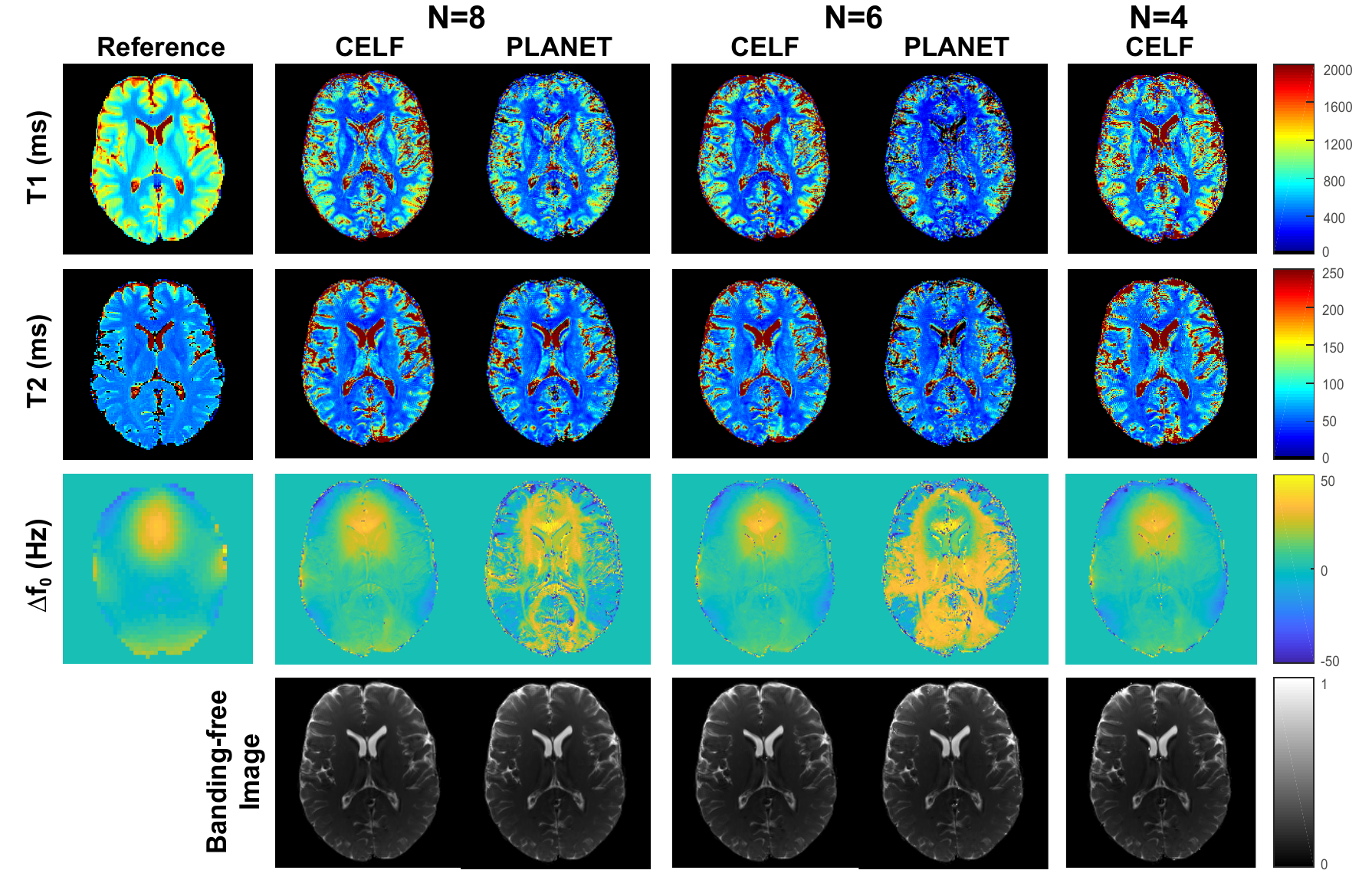}
\caption{\label{fig:invivo_jet_v3_subj1_supp} Parameter estimation was performed with CELF and PLANET on phase-cycled bSSFP images of the brain for S1. Results are shown in a representative cross-section for $N={\{4,6,8\}}$. $T_1$, $T_2$, and off-resonance estimates are displayed along with the reference maps from gold-standard mapping sequences, corresponding banding-free image estimates are provided.}
\end{figure*}

\begin{figure*}[t]
\centering
\includegraphics[width=\textwidth]{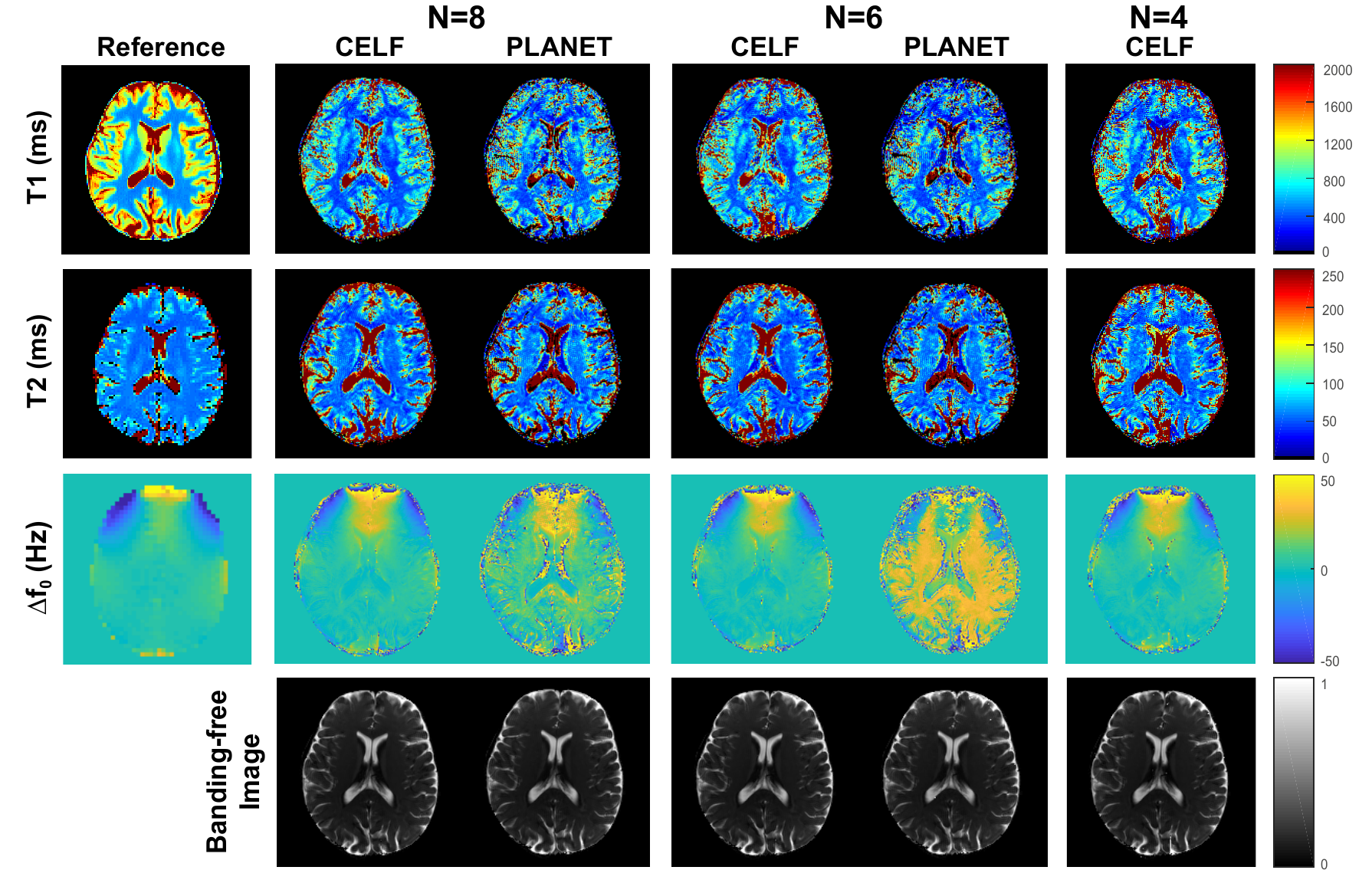}
\caption{\label{fig:invivo_jet_v3_subj2_supp} Parameter estimation was performed with CELF and PLANET on phase-cycled bSSFP images of the brain for S2. Results are shown in a representative cross-section for $N={\{4,6,8\}}$. $T_1$, $T_2$, and off-resonance estimates are displayed along with the reference maps from gold-standard mapping sequences, corresponding banding-free image estimates are provided.}
\end{figure*}

\begin{figure*}[t]
\centering
\includegraphics[width=0.85\columnwidth]{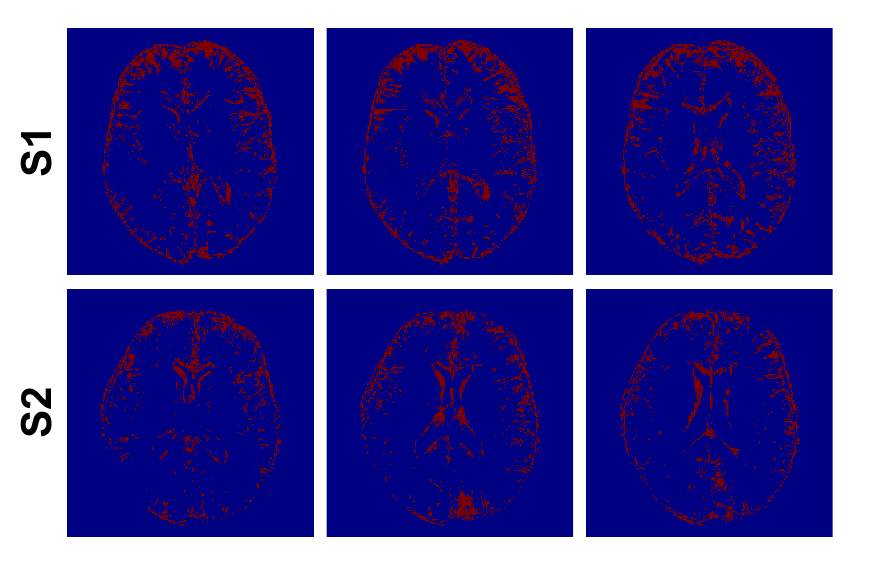}
\caption{\label{fig:invivo_subj12_gamma_supp} Brain voxels in which $\gamma$ is identified via the bounded search procedure as opposed to the analytical solution are marked in red color. Results are shown across three representative slices for subject S1 (top row) and for subject S2 (bottom row). Percentages of red voxels within the brain are 17.9$\%$ for S1 and 12.6$\%$ for S2.}
\end{figure*}

\begin{figure*}[t]
\centering
\includegraphics[width=\columnwidth]{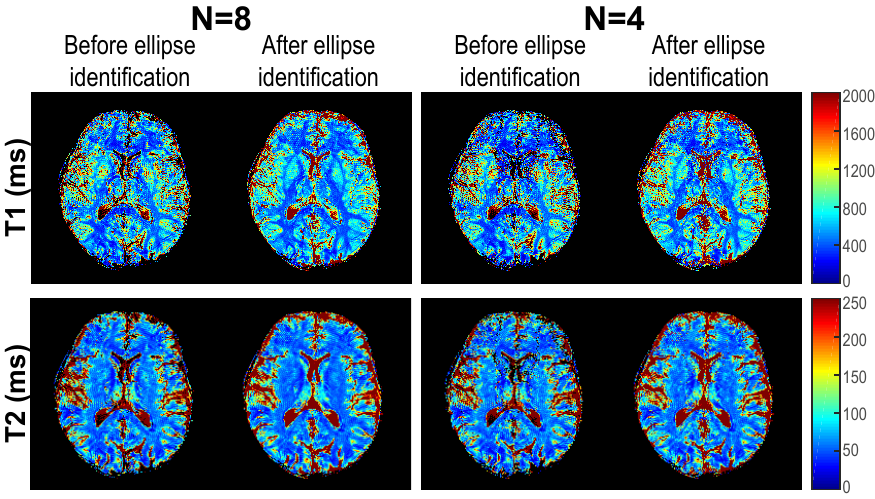}
\caption{\label{fig:invivo_celf} Parameter estimation was performed with CELF on phase-cycled bSSFP images of the brain. $T_1$ and $T_2$ estimates are shown in a representative cross-section in S2 for $N={\{4,8\}}$. Results are shown based on fit ellipses prior to and following dictionary-based ellipse identification.}
\end{figure*}

\begin{figure*}[t]
\centering
\includegraphics[width=\textwidth]{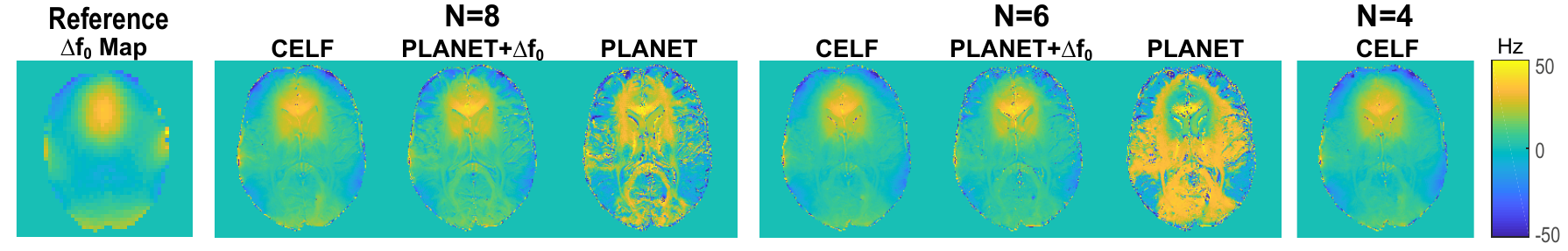}
\caption{\label{fig:invivo_B0_subj1_supp} Off-resonance map estimates from in vivo bSSFP brain images of S1 at $N=\{4,6,8\}$. Estimates are shown for CELF, PLANET and PLANET+$\Delta f_0$, a variant of PLANET modified to include the off-resonance estimation step in CELF. A reference map from a standard mapping sequence is also shown.}
\end{figure*}

\begin{figure*}[t]
\centering
\includegraphics[width=\textwidth]{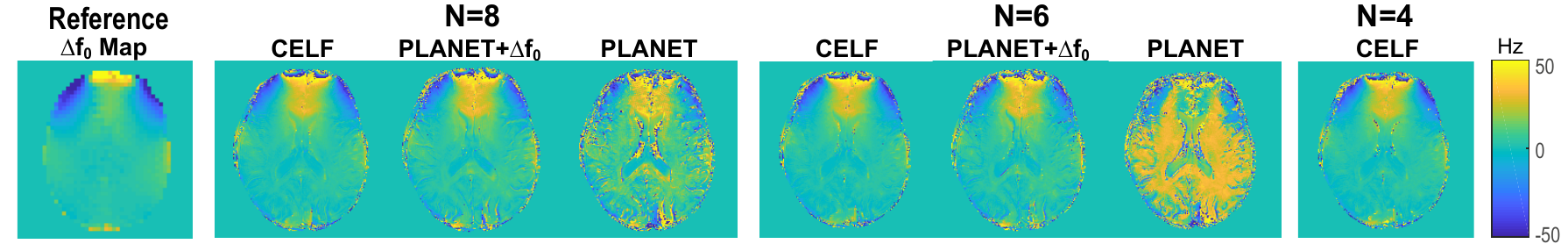}
\caption{\label{fig:invivo_B0_subj2_supp} Off-resonance map estimates from in vivo bSSFP brain images of S2 at $N=\{4,6,8\}$. Estimates are shown for CELF, PLANET and PLANET+$\Delta f_0$, a variant of PLANET modified to include the off-resonance estimation step in CELF. A reference map from a standard mapping sequence is also shown.}
\end{figure*}

\begin{figure*}[t]
\centering
\includegraphics[width=0.7\columnwidth]{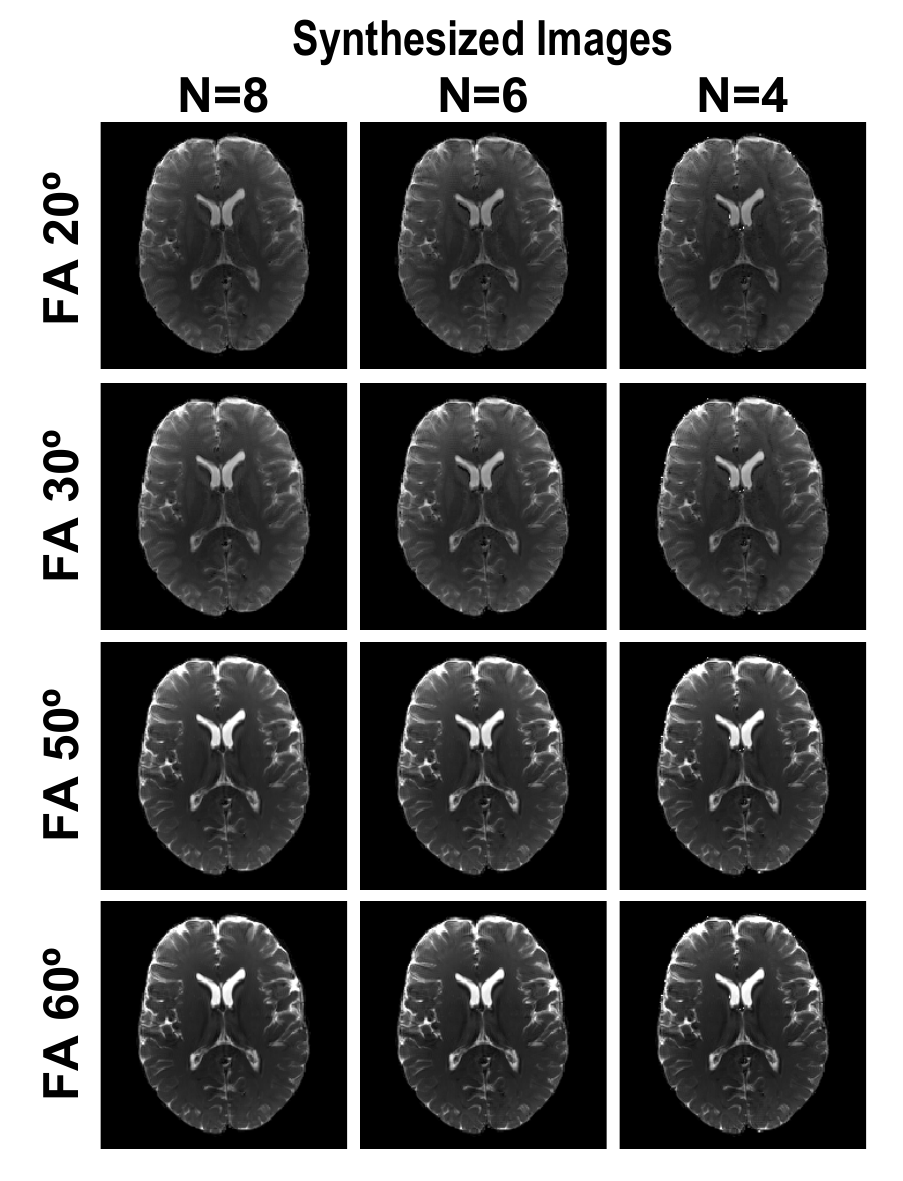}
\caption{\label{fig:invivo_subj1_FAsynt_supp} Synthetic bSSFP images at varying flip angles were generated based on CELF parameter estimates obtained a a specific flip angle in S1. First, $T_1$, $T_2$, and banding free image estimates at flip angle $=40^o$ were obtained with CELF. Next, CELF parameter estimates were used to synthesize banding-free bSSFP images at flip angles $\{20^o, 30^o, 50^o, 60^o\}$. As expected, CSF appears brighter towards higher flip angles, whereas gray and white-matter signals diminish.}
\end{figure*}

\begin{figure*}[t]
\centering
\includegraphics[width=0.7\columnwidth]{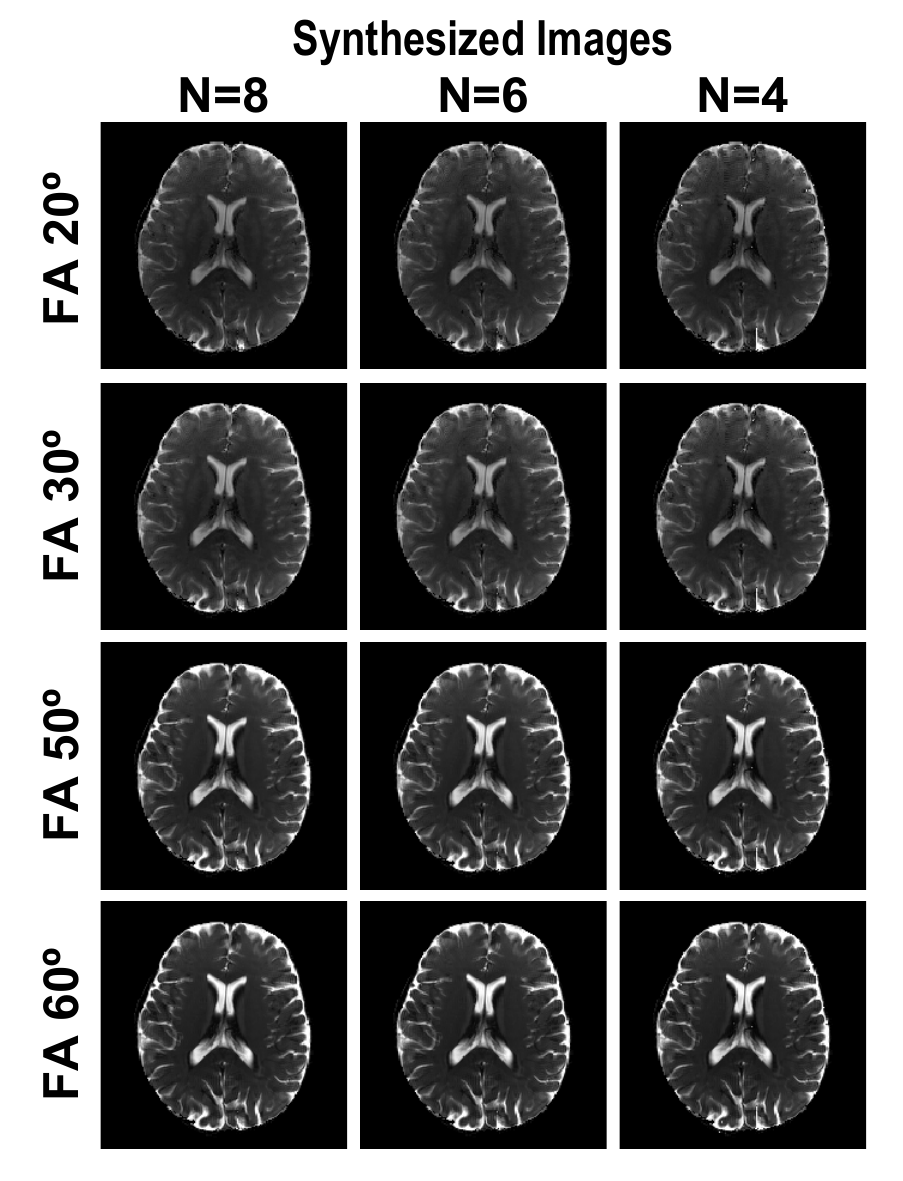}
\caption{\label{fig:invivo_subj2_FAsynt_supp} Synthetic bSSFP images at varying flip angles were generated based on CELF parameter estimates obtained a a specific flip angle in S2. First, $T_1$, $T_2$, and banding free image estimates at flip angle $=40^o$ were obtained with CELF. Next, CELF parameter estimates were used to synthesize banding-free bSSFP images at flip angles $\{20^o, 30^o, 50^o, 60^o\}$. As expected, CSF appears brighter towards higher flip angles, whereas gray and white-matter signals diminish.}
\end{figure*}

\begin{figure*}[t]
\centering
\includegraphics[width=\columnwidth]{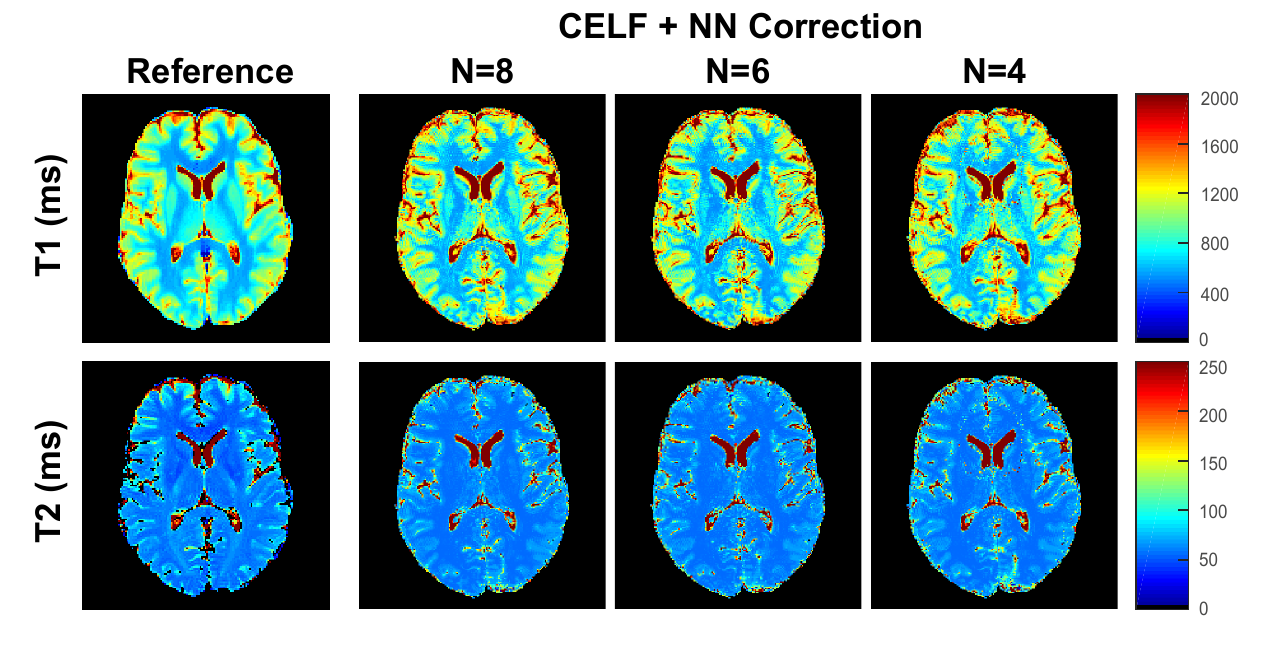}
\caption{\label{fig:invivo_NN_subj1_supp} A learning-based correction procedure was performed on CELF-derived parameter estimates. Results are shown in a representative cross-section from Subject 1 for $N={\{4,6,8\}}$. $T_1$ and $T_2$ estimates are displayed along with the reference maps from gold-standard mapping sequences.}
\end{figure*}

\clearpage
\printbibliography[heading=subbibliography] 
\end{refsection}




\end{document}